\documentclass[letterpaper,11pt]{article}

\usepackage{bm}
\usepackage{url}
\usepackage{array}
\usepackage{dsfont}
\usepackage{amsmath}
\usepackage{amsthm}
\usepackage{amssymb}
\usepackage{caption}
\usepackage{amsfonts}
\usepackage{booktabs}
\usepackage{graphicx}
\usepackage{mathrsfs}
\usepackage{multicol}
\usepackage{multirow}
\usepackage{cellspace}
\usepackage{enumerate}
\usepackage{enumitem}
\usepackage{makecell}
\usepackage{mathtools}
\usepackage[dvipsnames]{xcolor}
\usepackage{tikz}
\usetikzlibrary{arrows.meta,calc,positioning,backgrounds}

\usepackage{hyperref}
\hypersetup{
	linktoc      = all,
	colorlinks   = true,
	urlcolor     = blue,
	linkcolor    = blue,
	citecolor    = red
}

\usepackage[backend=biber, isbn=false, style=alphabetic, backref=true, doi=true, url=true, maxcitenames=10, mincitenames=10, maxalphanames=10, maxbibnames=10, minbibnames=5, minalphanames=5, defernumbers=true, sortlocale=en_US]{biblatex}
\usepackage[T1]{fontenc}
\usepackage[margin=1in]{geometry}
\usepackage[capitalize,noabbrev]{cleveref}
\usepackage[caption=false,font=normalsize,labelfont=sf,textfont=sf]{subfig}

\usepackage{titlesec}
\titlespacing*{\paragraph}{0pt}{1ex}{1ex}

\setlist[itemize]{leftmargin=*}
\setlist[enumerate]{label=(\arabic*),leftmargin=*}

\newtheorem{theorem}{Theorem}[section]
\newtheorem{lemma}[theorem]{Lemma}

\crefname{lemma}{Lemma}{Lemmas}
\Crefname{lemma}{Lemma}{Lemmas}

\DeclareMathOperator{\supp}{supp}
\DeclareMathOperator{\E}{\mathbb{E}}

\DeclareMathOperator{\rank}{rank}
\DeclareMathOperator{\range}{range}
\DeclareMathOperator{\tr}{tr}
\DeclareMathOperator{\TV}{TV}
\DeclareMathOperator{\lev}{lev}
\DeclareMathOperator{\diag}{diag}

\def\eps{\varepsilon}

\def\R{\mathbb{R}}

\def\tO{\widetilde{O}}

\def\mat{\mathbf}

\def\mA{\mat{A}}
\def\mB{\mat{B}}

\def\mI{\mat{I}}
\def\mL{\mat{L}}
\def\mM{\mat{M}}
\def\mN{\mat{N}}

\def\mT{\mat{T}}
\def\mU{\mat{U}}
\def\mW{\mat{W}}

\def\mzero{\mat{0}}

\def\vec{\boldsymbol}

\def\va{\vec{a}}
\def\vb{\vec{b}}
\def\vc{\vec{c}}
\def\ve{\vec{e}}
\def\vf{\vec{f}}

\def\vu{\vec{u}}
\def\vv{\vec{v}}
\def\vw{\vec{w}}
\def\vx{\vec{x}}

\def\tw{\widetilde{w}}
\def\tvw{\widetilde{\vw}}

\def\valpha{\vec{\alpha}}
\def\vbeta{\vec{\beta}}
\def\vphi{\vec{\phi}}

\def\vzero{\vec{0}}

\def\set{\mathcal}

\def\sB{\set{B}}
\def\sC{\set{C}}

\def\sI{\set{I}}
\def\sR{\set{R}}
\def\sS{\set{S}}
\def\sT{\set{T}}

\def\Yes{\mathsf{Yes}}
\def\No{\mathsf{No}}

\def\event{\mathcal{E}}

\title{A Tight Superlinear Lower Bound on the Sketching Dimension of Weighted Spectral Sparsification}

\author{
	Yanyu Chen\thanks{National University of Singapore. Email: \texttt{yanyu.chen@u.nus.edu}.}
	\and Yu Chen\thanks{National University of Singapore. Email: \texttt{yu.chen@nus.edu.sg}.}
	\and Mingji Yang\thanks{Renmin University of China. Email: \texttt{kyleyoung@ruc.edu.cn}.}
}

\date{}

\begin{document}

\maketitle

\begin{abstract}

We show that any linear sketch that computes a $\bigl(1 \pm O(1)\bigr)$-spectral sparsifier of $n$-vertex weighted graphs with polynomially bounded weights must have dimension $\Omega\bigl(n^{6/5}\bigr)$.
Previously, Chen, Khanna, and Li (FOCS 2022) gave a sketching dimension upper bound of $\tO\bigl(n^{6/5}\bigr)$ and a lower bound of $\Omega\bigl(n^{21/20 - o(1)}\bigr)$ for incidence sketches, in which each measurement examines edges incident on a single vertex.
Our lower bound removes this restriction and matches the upper bound up to polylogarithmic factors.
Together with their $\tO(n)$ upper bound for weighted cut sparsification, our result also gives a sharp separation between cut and spectral sparsification.

In our hard distributions, we randomly permute vertices into blocks on a cycle and construct a complete graph whose edge weights decrease with cyclic block distance.
We reduce the problem to showing that the sketch column of an edge between opposite blocks has a small leverage score relative to other columns scaled by the edge weights.
For incidence sketches, we obtain this bound from a routing decomposition on the block cycle that uses few columns between distant blocks.
We then add a separate clique to reduce arbitrary sketches to incidence sketches with expected dimension linear in the original dimension.

\end{abstract}

\section{Introduction} \label{sec:intro}

Graph sparsification seeks to compress a graph into a sparse graph that approximately preserves some properties of the original graph.
It is a fundamental tool in graph algorithms, with applications to approximating minimum cuts and maximum flows~\cite{She13nearly,KLOS14almost,BK15randomized,Peng16approximate}, solving Laplacian systems~\cite{ST14nearly,JS25ultrasparse}, and clustering graphs~\cite{CSWZ16communication}.
Two important notions of graph sparsification are \textit{spectral sparsification}~\cite{ST11spectral}, which approximately preserves every quadratic form of the graph Laplacian, and \textit{cut sparsification}~\cite{BK98approximating}, which approximately preserves the weight of every cut.

We first introduce the definitions of spectral and cut sparsification.
For $\vw \in \R_{\ge 0}^{\binom{n}{2}}$, the Laplacian of the weighted graph with weight vector $\vw$ is
\begin{align*}
	\mL_{\vw} \coloneqq \sum_{uv \, \in \, \binom{[n]}{2}} w_{uv} (\ve_u-\ve_v)(\ve_u-\ve_v)^{\top}.
\end{align*}
We call a symmetric matrix $\widehat{\mL}$ a $(1 \pm \eps)$-\textit{spectral sparsifier} of $\mL_{\vw}$ if
\begin{align*}
	(1-\eps) \vx^{\top} \mL_{\vw} \vx \le \vx^{\top} \widehat{\mL} \, \vx \le (1+\eps) \vx^{\top} \mL_{\vw} \vx \qquad \text{for all } \vx \in \R^n.
\end{align*}
\textit{Cut sparsification} only requires this inequality for $\vx \in \{0,1\}^n$.
Thus, every spectral sparsifier is also a cut sparsifier.
Despite the stronger requirement of spectral sparsification, both notions admit sparsifiers of nearly linear size for constant approximation error.
Spielman and Srivastava~\cite{SS11graph} showed how to construct a $(1 \pm \eps)$-spectral sparsifier with $O\bigl(n \log n / \eps^2\bigr)$ edges in nearly linear time, and Batson, Spielman, and Srivastava~\cite{BSS12twice} proved the existence of such a sparsifier with $O\bigl(n / \eps^2\bigr)$ edges.

The two forms of sparsification also have similar complexity bounds in many computational settings, including sequential constructions~\cite{BK15randomized,SS11graph}, parallel and distributed algorithms~\cite{KX16simple}, and fully dynamic algorithms~\cite{ADKKP16fully}.
These similar bounds suggest a close connection between the complexity of cut and spectral sparsification in many settings.

We show, however, that spectral sparsification requires polynomially more measurements than cut sparsification in the \textit{linear sketching} model for weighted graphs.
Specifically, for a fixed constant approximation error, $\tO(n)$ measurements suffice for cut sparsification~\cite{CKL22weighted}, whereas spectral sparsification requires $\Omega\bigl(n^{6/5}\bigr)$ measurements, even for polynomially bounded weight ratios.
In this model, a randomized linear sketch of dimension $N$ samples a matrix $\mA \in \R^{N \times \binom{n}{2}}$ independently of the input edge-weight vector $\vw$.
A recovery algorithm then computes a sparsifier from $\mA$ and the $N$ measurements $\mA\vw$.

One motivation for this model comes from streaming algorithms, where a graph arrives as a sequence of edge-weight updates and the goal is to compute a sparsifier in one pass using limited space.
Linear sketches are well suited to \textit{turnstile streams}, which allow both increases and decreases in edge weights.
For each update, we multiply the weight update vector by the sketching matrix and add the result to the stored measurement vector, without storing the graph itself.
Conversely, under suitable assumptions on stream length and coordinate ranges, general one-pass turnstile algorithms can be simulated by linear sketches with comparable space complexity~\cite{LNW14turnstile,AHLW16new,KP20separations,JLY26turnstile}.
This connection makes linear sketching a natural starting point for proving streaming space lower bounds.

For unweighted graphs, linear sketches achieve nearly linear dimension for both cut and spectral sparsification.
Ahn, Guha, and McGregor~\cite{AGM12graph} initiated the study of graph sparsification by linear sketching and gave sketches for cut sparsification.
Kapralov, Lee, Musco, Musco, and Sidford~\cite{KLMMS17single} subsequently obtained spectral sparsifiers using sketches with $\tO\bigl(n / \eps^2\bigr)$ dimension, with later work improving the recovery time~\cite{KMMMNST20fast}.
For constant $\eps$, these results show that $\tO(n)$ linear measurements suffice to recover a sparsifier with $\tO(n)$ edges for turnstile streams.

For weighted graphs, nearly-linear-space algorithms for spectral sparsification in insertion-only streams have long been known~\cite{KL13spectral}, with more recent work improving the logarithmic factors~\cite{CAWXZ26nearly}.
For streams that allow insertions and deletions of weighted edges, grouping edges by weight also extends unweighted sketches to weighted graphs, provided that each deletion specifies the same weight as the corresponding insertion~\cite{CKL22weighted}.
However, general turnstile updates specify changes in edge weights, which may move an edge between groups.
Thus, updating the group sketches requires knowing the edge's current weight, so this approach does not directly give a small-space turnstile algorithm for weighted sparsification.
Grouping by weight is also nonlinear in $\vw$, so it does not directly yield a linear sketch for weighted cut or spectral sparsification.

The question is therefore whether the nearly-linear sketching dimension bounds for cut and spectral sparsification still hold for weighted graphs.
Chen, Khanna, and Li~\cite{CKL22weighted} studied this question and obtained an upper bound of $\tO\bigl(n / \eps^3\bigr)$ on the sketching dimension for weighted cut sparsification, which was recently improved to $\tO\bigl(n / \eps^2\bigr)$ by Lin, Mirrokni, and Woodruff~\cite{LMW26near}.
For weighted spectral sparsification, \cite{CKL22weighted} gave a sketching dimension upper bound of $\tO\bigl(n^{6/5} / \eps^4\bigr)$, with a polylogarithmic dependence on $w_{\max} / w_{\min}$, the ratio of the largest and smallest nonzero edge weights allowed in the input graph.
Their spectral algorithm combines vertex sampling with an analysis of effective resistances in vertex-induced subgraphs to recover spectrally important edges.

On the other hand, \cite{CKL22weighted} further proved a dimension lower bound of $\Omega\bigl(n^{21/20 - o(1)}\bigr)$ for weighted spectral sparsification by \textit{incidence sketches}, a natural but restricted class of sketches in which each measurement is supported on edges incident to a single vertex.
This lower bound established a separation within that class, but left two questions open for weighted spectral sparsification: whether arbitrary linear sketches require superlinear dimension, and whether the exponent $6/5$ in the upper bound is optimal.

\subsection{Our Results}

We resolve both questions by proving an $\Omega\bigl(n^{6/5}\bigr)$ lower bound for arbitrary linear sketches.
The bound holds even when the ratio between the largest and smallest nonzero edge weights is $O(n^5)$.

\begin{theorem} \label{thm:main}
	Any linear sketch of dimension $N$ that computes a $(1 \pm 0.01)$-spectral sparsifier with probability at least $0.99$ on every $n$-vertex graph with $w_{\max} / w_{\min} = O\bigl(n^5\bigr)$ must satisfy $N = \Omega\bigl(n^{6/5}\bigr)$.
\end{theorem}

For constant approximation error and polynomially bounded weight ratio, \cref{thm:main} matches the upper bound of \cite{CKL22weighted} up to polylogarithmic factors.
It improves the exponent of their lower bound and removes the restriction to incidence sketches.
This result also holds when the recovery algorithm may output any symmetric matrix satisfying the spectral approximation guarantee, without a sparsity requirement on the output.

Together with the $\tO(n)$ upper bound for weighted cut sparsification in~\cite{CKL22weighted}, our result gives a sharp separation between cut and spectral sparsification.
Requiring a spectral sparsifier increases the necessary sketching dimension from nearly linear to $\Omega\bigl(n^{6/5}\bigr)$.
The $\tO(n)$ upper bound for unweighted spectral sparsification~\cite{KLMMS17single} further shows that edge weights are essential to this lower bound.

A different separation between cut and spectral sparsification was obtained by Chen, Shi, and Trevisan~\cite{CST22cut}, who compared the approximation error achievable by weighted subgraph sparsifiers with a prescribed number of edges.
Our separation concerns the number of linear measurements needed to compute a sparsifier.

\paragraph{Connection to turnstile streaming.}
Our lower bound also suggests a route toward space lower bounds for general one-pass turnstile streaming algorithms.
A line of work shows how to simulate general streaming algorithms by linear sketches under various assumptions on the stream~\cite{LNW14turnstile,AHLW16new,JLY26turnstile}.
Together with techniques for transferring real-valued sketching lower bounds to discrete inputs~\cite{GLWYZ25lifting}, these simulations could extend dimension lower bounds such as ours to streaming space lower bounds.
Such transfers depend on the allowed stream length and coordinate range~\cite{KP20separations}.
Adapting these techniques to our hard distributions and identifying the precise streaming settings in which the resulting lower bound holds remain open problems.

\section{Technical Overview} \label{sec:overview}

We use a standard framework to establish the lower bound: construct two ``hard'' input distributions $\Yes$ and $\No$ that can be distinguished using a spectral sparsifier, and then show that the corresponding output distributions by linear sketching have small total variation distance (TV distance) if the sketch has too few rows.
We first prove the lower bound for incidence sketches, in which each row is supported on edges incident to a single vertex.

\paragraph{The hard distributions.}
Inspired by the construction in \cite{CKL22weighted}, we randomly permute the vertices into equally sized blocks $\sB_0,\dots,\sB_{\ell-1}$ arranged on a cycle, as illustrated in \cref{fig:overview-ring-structure}.
The graph is complete, and edges within a block have large base weights, while weights between distinct blocks decay as the inverse fourth power of their cyclic distance.
Under $\No$, we multiply each base weight $w_e$ by independent noise supported on $[1/2,3/2]$; under $\Yes$, we additionally increase the weight of a query edge $q = xy$ joining $\sB_0$ and $\sB_{\ell/2}$ by one.
A potential that is constant within blocks and varies gradually around the cycle reveals an energy gap that a sufficiently accurate spectral sparsifier must detect.

\definecolor{ovInk}{HTML}{263442}
\definecolor{ovMuted}{HTML}{7D8996}
\definecolor{ovBlue}{HTML}{0072B2}
\definecolor{ovPurple}{HTML}{8B5AA5}
\definecolor{ovOrange}{HTML}{D78E59}
\definecolor{ovDeepOrange}{HTML}{8F321D}
\definecolor{ovTeal}{HTML}{008B78}
\definecolor{ovFill}{HTML}{F3F5F7}
\tikzset{
  ov/base/.style={font=\small, text=ovInk, line cap=round, line join=round},
  ov/block/.style={draw=ovMuted!75, fill=white, rounded corners=2pt,
    line width=.55pt, minimum width=.78cm, minimum height=.62cm, inner sep=2pt},
  ov/region/.style={draw=ovMuted!65, fill=ovFill, rounded corners=5pt,
    line width=.55pt},
  ov/analysis block/.style={ov/region,draw=ovTeal,fill=ovTeal!3,
    line width=.7pt},
  ov/slot/.style={draw=ovMuted!55, fill=white, rounded corners=3pt,
    line width=.45pt},
  ov/vertex/.style={circle, draw=none, fill=ovInk,
    minimum size=3.2pt, inner sep=0pt},
  ov/endpoint/.style={circle, draw=none, fill=ovBlue,
    minimum size=4.4pt, inner sep=0pt},
  ov/query/.style={draw=ovBlue, line width=1.2pt,
    dash pattern=on 4.5pt off 2.5pt},
  ov/root/.style={draw=ovPurple, line width=1.2pt},
  ov/root faint/.style={draw=ovPurple!14, line width=.55pt},
  ov/source/.style={draw=ovPurple, line width=1.0pt,
    dash pattern=on 4.5pt off 2.5pt},
  ov/source faint/.style={draw=ovPurple!14, line width=.55pt,
    dash pattern=on 4.5pt off 2.5pt},
  ov/local/.style={draw=ovOrange, line width=.9pt},
  ov/local faint/.style={draw=ovOrange!12, line width=.45pt},
  ov/cross/.style={draw=ovDeepOrange, line width=1.2pt},
  ov/cross faint/.style={draw=ovDeepOrange!10, line width=.45pt},
  ov/star/.style={draw=ovTeal, line width=.95pt,
    dash pattern=on 1.2pt off 1.7pt},
  ov/star faint/.style={draw=ovTeal!14, line width=.55pt,
    dash pattern=on 1.2pt off 1.7pt},
  ov/route/.style={draw=ovTeal, line width=1.25pt,
    -{Stealth[length=4.5pt,width=4pt]}},
  ov/note/.style={font=\footnotesize, text=ovMuted, align=center},
  ov/key/.style={font=\footnotesize, anchor=west},
  ov/label/.style={fill=none, inner sep=2pt, font=\small},
}

\newcommand{\OverviewTransitionBlock}[5][0]{%
  \begin{scope}[shift={(#4,#5)}]
    \draw[ov/analysis block] (0,-#1) rectangle (6.10,1.97);
    \node[anchor=north east,inner sep=0pt,outer sep=0pt,xshift=-3pt]
      at (0,1.97) {$\sB_{#3}$};
    \draw[ov/slot] (.22,.18) rectangle (2.33,1.78);
    \draw[ov/slot] (2.53,.18) rectangle (5.88,1.78);
    \coordinate (#2-reserve-label) at (1.28,1.47);
    \coordinate (#2-auxiliary-label) at (4.21,1.47);
    \coordinate (#2-t1) at (.67,1.20);
    \coordinate (#2-t2) at (1.89,1.20);
    \coordinate (#2-t3) at (.67,.50);
    \coordinate (#2-t4) at (1.89,.50);
    \coordinate (#2-s1) at (3.05,1.12);
    \coordinate (#2-s2) at (4.20,1.12);
    \coordinate (#2-s3) at (5.35,1.12);
    \coordinate (#2-s4) at (3.05,.42);
    \coordinate (#2-s5) at (4.20,.42);
    \coordinate (#2-s6) at (5.35,.42);
    \coordinate (#2-gate1) at (.40,.02);
    \coordinate (#2-gate2) at (2.16,.02);
    \coordinate (#2-gate3) at (.86,.02);
    \coordinate (#2-gate4) at (1.68,.02);
    \coordinate (#2-query) at (.85,-.80);
    \coordinate (#2-route-out) at (4.50,1.97);
    \coordinate (#2-route-in) at (1.60,1.97);
  \end{scope}%
}

\newcommand{\OverviewTransitionClique}[2]{%
  \foreach \i in {1,...,4} {
    \foreach \j in {1,...,4} {
      \ifnum\j>\i \draw[ov/local faint] (#1-t\i)--(#1-t\j);\fi
    }
  }
  \ifcase#2\relax
  \or
    \draw[ov/local] (#1-t1)--(#1-t2)--(#1-t4)--(#1-t3);
  \or
    \draw[ov/local] (#1-t1)--(#1-t3)--(#1-t4)--(#1-t1);
  \or
    \draw[ov/local] (#1-t1)--(#1-t2)--(#1-t3)--(#1-t1);
  \or
    \draw[ov/local] (#1-t1)--(#1-t4)--(#1-t2)--(#1-t3);
  \fi
}

\newcommand{\OverviewTransitionArrow}[2]{%
  \draw[ov/route] (#1-route-out)
    .. controls ($(#1-route-out)+(.40,.95)$)
       and ($(#2-route-in)+(-.45,.95)$)
    .. (#2-route-in);
}

\newcommand{\OverviewTransitionBridge}[3]{%
  \foreach \i/\j in {1/1,2/2,3/3,4/2,5/4,6/1} {
    \draw[ov/local faint] (#1-s\i)--(#1-t\j);
  }
  \foreach \i/\j in {1/2,2/3,3/1,4/4,5/2,6/3} {
    \draw[ov/cross faint] (#1-s\i)--(#2-t\j);
  }
  \ifnum#3=1
    \draw[ov/local] (#1-t2)--(#1-s1) (#1-t4)--(#1-s3);
    \draw[ov/cross] (#1-s3)--(#2-t3) (#1-s6)--(#2-t2);
  \else
    \draw[ov/local] (#1-t1)--(#1-s4) (#1-t4)--(#1-s2);
    \draw[ov/cross] (#1-s2)--(#2-t1) (#1-s5)--(#2-t4);
  \fi
}

\newcommand{\OverviewTransitionRootColumns}[3]{%
  \foreach \i/\dx/\depth in {1/-.40/1.10,2/.40/.70,3/0/1.30,4/0/.90} {
    \draw[ov/source faint] (#2-t\i)
      .. controls ($(#2-t\i)+(\dx,-.15)$) and ($(#2-gate\i)+(0,.45)$)
      .. (#2-gate\i)
      .. controls ($(#2-gate\i)+(1.20,-\depth)$) and ($(#1)+(-1.80,-.10)$)
      .. (#1);
    \draw[ov/source faint] (#1)--(#3-t\i);
  }
  \foreach \i/\dx/\depth in {2/.40/.70,3/0/1.30} {
    \draw[ov/source] (#2-t\i)
      .. controls ($(#2-t\i)+(\dx,-.15)$) and ($(#2-gate\i)+(0,.45)$)
      .. (#2-gate\i)
      .. controls ($(#2-gate\i)+(1.20,-\depth)$) and ($(#1)+(-1.80,-.10)$)
      .. (#1);
  }
  \draw[ov/source] (#1)--(#3-t1) (#1)--(#3-t4);
}

\newcommand{\OverviewTransitionLegend}{%
  \begin{scope}[x=1cm]
    \draw[ov/source] (-6.40,-1.95)--(-5.85,-1.95);
    \node[ov/key] at (-5.70,-1.95) {root-star terms};
    \draw[ov/local] (-2.45,-1.95)--(-1.90,-1.95);
    \node[ov/key] at (-1.75,-1.95) {within-block};
    \draw[ov/cross] (.85,-1.95)--(1.40,-1.95);
    \node[ov/key] at (1.55,-1.95) {cross-block};
    \draw[ov/route] (4.10,-1.95)--(4.70,-1.95);
    \node[ov/key] at (4.85,-1.95) {transition};
  \end{scope}%
}

\newcommand{\OverviewTransitionVertices}[3]{%
  \foreach \i in {1,...,4} {\node[ov/vertex] at (#1-t\i) {};}
  \foreach \i in {1,...,6} {\node[ov/vertex,fill=#2] at (#1-s\i) {};}
  \node[font=\footnotesize,inner sep=1pt]
    at (#1-reserve-label) {$\sT_{#3}$};
  \node[font=\footnotesize,inner sep=1pt]
    at (#1-auxiliary-label) {$\sS_{#3}$};
}

\newcommand{\OverviewFigure}[3]{%
  \par\addvspace{8pt}\noindent
  \begin{minipage}{\linewidth}
    \centering
    \captionsetup{type=figure,font=small,skip=4pt}
    \scalebox{.92}{\input{#1}}
    \caption{#2}\label{#3}
  \end{minipage}
  \par\addvspace{8pt}
}

\newcommand{\OverviewFigurePair}[8]{%
  \par\addvspace{8pt}\noindent
  \begin{minipage}{\linewidth}
    \centering
    \captionsetup{type=figure,font=small,skip=4pt}
    \captionsetup[subfloat]{font=small,labelfont=rm,textfont=rm}
    \subfloat[#2\label{#3}]{%
      \begin{minipage}[b]{.48\linewidth}
        \centering
        \input{#1}
      \end{minipage}%
    }\hfill
    \subfloat[#5\label{#6}]{%
      \begin{minipage}[b]{.48\linewidth}
        \centering
        \input{#4}
      \end{minipage}%
    }
    \caption{#7}\label{#8}
  \end{minipage}
  \par\addvspace{8pt}
}

\newcommand{\OverviewRingFigure}{%
  \begingroup
  \captionsetup[subfloat]{farskip=0pt}%
  \par\addvspace{8pt}\noindent
  \begin{minipage}{\linewidth}
    \centering
    \captionsetup{type=figure,font=small,skip=4pt}
    \captionsetup[subfloat]{font=small,labelfont=rm,textfont=rm}
    \subfloat[Ring structure.\label{fig:overview-ring-structure}]{%
      \begin{minipage}[b]{.48\linewidth}
        \centering
        \begin{tikzpicture}[ov/base]
  \path[use as bounding box] (-3.75,-2.25) rectangle (3.75,2.25);
  \foreach \i in {0,...,7} {
    \coordinate (b\i) at ({2.32*cos(180-45*\i)},{1.85*sin(180-45*\i)});
  }
  \foreach \i in {0,...,7} {
    \foreach \j in {0,...,7} {
      \ifnum\j>\i
        \pgfmathtruncatemacro{\dist}{min(\j-\i,8-\j+\i)}
        \ifcase\dist\or
          \draw[ovMuted!85,line width=1.3pt] (b\i)--(b\j);
        \or
          \draw[ovMuted!60,line width=.65pt] (b\i)--(b\j);
        \or
          \draw[ovMuted!45,line width=.35pt] (b\i)--(b\j);
        \or
          \draw[ovMuted!35,line width=.2pt] (b\i)--(b\j);
        \fi
      \fi
    }
  }
  \node[ov/block] at (b1) {$\sB_1$};
  \node[ov/block] at (b2) {$\sB_2$};
  \node[ov/block] at (b6) {$\sB_{\ell-2}$};
  \node[ov/block] at (b7) {$\sB_{\ell-1}$};
  \foreach \i in {3,5} {
    \node[ov/block] at (b\i) {$\cdots$};
  }
  \node[ov/block,minimum width=1.10cm,minimum height=.92cm] at (b0) {};
  \node[ov/block,minimum width=1.10cm,minimum height=.92cm] at (b4) {};
  \node[font=\footnotesize] at ($(b0)+(0,.23)$) {$\sB_0$};
  \node[font=\footnotesize] at ($(b4)+(0,.23)$) {$\sB_{\ell/2}$};
  \coordinate (x) at ($(b0)+(.13,-.16)$);
  \coordinate (y) at ($(b4)+(-.13,-.16)$);
  \draw[ov/query] (x)--(y);
  \node[ov/endpoint] at (x) {};
  \node[ov/endpoint] at (y) {};
  \node[font=\footnotesize,text=ovBlue] at ($(b0)+(-.24,-.16)$) {$x$};
  \node[font=\footnotesize,text=ovBlue] at ($(b4)+(.24,-.16)$) {$y$};
  \node[ov/label,text=ovBlue] at (0,.13) {$q=xy$};
\end{tikzpicture}
      \end{minipage}%
    }\hfill
    \subfloat[The first representation.\label{fig:overview-reserve}]{%
      \begin{minipage}[b]{.48\linewidth}
        \centering
        \begin{tikzpicture}[ov/base]
  \path[use as bounding box] (-3.75,-2.25) rectangle (3.75,2.25);
  \draw[ov/analysis block] (-3.45,-1.18) rectangle (-.25,1.48);
  \node[anchor=north east,inner sep=0pt,outer sep=0pt,xshift=-3pt,
    font=\footnotesize] at (-3.45,1.48) {$\sB_0$};
  \draw[ov/slot] (-3.20,-.40) rectangle (-.50,1.35);
  \node at (-1.85,.96) {$\sT_0$};
  \coordinate (x) at (-2.65,-.82);
  \coordinate (y) at (3.00,-.82);
  \coordinate (t1) at (-2.65,.64);
  \coordinate (t2) at (-1.10,.64);
  \coordinate (t3) at (-2.65,-.12);
  \coordinate (t4) at (-1.10,-.12);
  \foreach \i in {1,...,4} {
    \draw[ov/local faint] (x)--(t\i);
    \draw[ov/root faint] (y)--(t\i);
    \foreach \j in {1,...,4} {
      \ifnum\j>\i \draw[ov/local faint] (t\i)--(t\j);\fi
    }
  }
  \draw[ov/local] (t1)--(t2) (t2)--(t4);
  \draw[ov/local] (x)--(t2) (x)--(t3);
  \draw[ov/root] (y)--(t2) (y)--(t4);
  \draw[ov/query] (x)--node[ov/label,text=ovBlue,pos=.58,below=3pt] {$\vb_{xy}$} (y);
  \node[ov/endpoint,label=left:$x$] at (x) {};
  \node[ov/endpoint,label=right:$y$] at (y) {};
  \foreach \i in {1,...,4} {\node[ov/vertex] at (t\i) {};}
\end{tikzpicture}
      \end{minipage}%
    }
    \caption{\textbf{The ring and the first representation.}
     (a) Every block is a clique and all block pairs are joined; thickness indicates the scale of the base weight.
     (b) For reserve-good $x$, we represent the dashed blue target $\vb_{xy}$ by orange within-block columns and purple $\sT_0$--$y$ columns (corresponding to the root-star terms).}\label{fig:overview-ring}
  \end{minipage}
  \par\addvspace{8pt}
  \endgroup
}

\newcommand{\OverviewTransitionFigure}{%
  \par\addvspace{8pt}\noindent
  \begin{minipage}{\linewidth}
    \centering
    \captionsetup{type=figure,font=small,skip=4pt}
    \scalebox{.92}{\begin{tikzpicture}[ov/base]
  \path[use as bounding box] (-7.10,-2.18) rectangle (7.25,2.98);
  \OverviewTransitionBlock[1.15]{pzero}{0}{-6.65}{-.15}
  \OverviewTransitionBlock{pone}{1}{.55}{.30}
  \OverviewTransitionBridge{pzero}{pone}{1}
  \OverviewTransitionClique{pzero}{1}
  \OverviewTransitionClique{pone}{2}
  \OverviewTransitionArrow{pzero}{pone}
  \coordinate (root) at (1.83,-.95);
  \OverviewTransitionRootColumns{root}{pzero}{pone}
  \OverviewTransitionVertices{pzero}{ovInk}{0}
  \OverviewTransitionVertices{pone}{ovMuted!55}{1}
  \node[ov/endpoint,label=left:$x$] at (pzero-query) {};
  \node[ov/endpoint,label=right:$y$] at (root) {};
  \OverviewTransitionLegend
\end{tikzpicture}}
    \caption{\textbf{The transition from $\sB_0$ to $\sB_1$.}}\label{fig:overview-transition}
  \end{minipage}
  \par\addvspace{8pt}
}

\newcommand{\OverviewSecondTransitionFigure}{%
  \par\addvspace{8pt}\noindent
  \begin{minipage}{\linewidth}
    \centering
    \captionsetup{type=figure,font=small,skip=4pt}
    \scalebox{.92}{\begin{tikzpicture}[ov/base,x=.70cm]
  \path[use as bounding box] (-10.55,-2.18) rectangle (10.90,3.43);
  \OverviewTransitionBlock[1.15]{pzero}{0}{-10.05}{-.15}
  \OverviewTransitionBlock{pone}{1}{-2.85}{.30}
  \OverviewTransitionBlock{ptwo}{2}{4.35}{.75}
  \OverviewTransitionBridge{pzero}{pone}{1}
  \OverviewTransitionBridge{pone}{ptwo}{2}
  \OverviewTransitionClique{pzero}{1}
  \OverviewTransitionClique{pone}{3}
  \OverviewTransitionClique{ptwo}{4}
  \OverviewTransitionArrow{pzero}{pone}
  \OverviewTransitionArrow{pone}{ptwo}
  \coordinate (root) at (5.63,-.95);
  \OverviewTransitionRootColumns{root}{pzero}{ptwo}
  \OverviewTransitionVertices{pzero}{ovInk}{0}
  \OverviewTransitionVertices{pone}{ovInk}{1}
  \OverviewTransitionVertices{ptwo}{ovMuted!55}{2}
  \node[ov/endpoint,label=left:$x$] at (pzero-query) {};
  \node[ov/endpoint,label=right:$y$] at (root) {};
  \OverviewTransitionLegend
\end{tikzpicture}}
    \caption{\textbf{Continuing the transition to $\sB_2$.} The root-star terms between $y$ and $\sB_1$ are cancelled.}\label{fig:overview-second-transition}
  \end{minipage}
  \par\addvspace{8pt}
}

\newcommand{\OverviewRoutingFigure}{%
  \par\addvspace{8pt}\noindent
  \begin{minipage}{\linewidth}
    \centering
    \captionsetup{type=figure,font=small,skip=4pt}
    \scalebox{.92}{\begin{tikzpicture}[ov/base]
  \path[use as bounding box] (-5.10,-3.45) rectangle (5.45,3.5);
  
  \foreach \i in {0,...,15} {
    \coordinate (r\i) at ({3.80*cos(180-22.5*\i)},{2.65*sin(180-22.5*\i)});
  }
  \foreach \i in {0,...,15} {
    \foreach \j in {0,...,15} {
      \ifnum\j>\i
        \pgfmathtruncatemacro{\dist}{min(\j-\i,16-\j+\i)}
        \ifnum\dist=1
          \draw[ovMuted!55,line width=.85pt] (r\i)--(r\j);
        \else
          \draw[ovMuted!11,line width=.18pt] (r\i)--(r\j);
        \fi
      \fi
    }
  }
  
  \foreach \i/\j in {0/1,1/2,4/5,5/6} {
    \draw[ov/cross] (r\i)--(r\j);
  }
  \draw[ov/cross] (r2) to[bend right=10] (r4);
  \draw[ov/cross] (r6) to[bend right=10] (r8);
  \foreach \i in {3,7,9,10,11,12,13} {
    \node[ov/block,draw=ovMuted!45,text=ovMuted,
      minimum width=.63cm,minimum height=.46cm,font=\footnotesize] at (r\i) {$\cdots$};
  }
  \node[ov/block,text=ovMuted,font=\footnotesize] at (r14) {$\sB_{\ell-2}$};
  \node[ov/block,text=ovMuted,font=\footnotesize] at (r15) {$\sB_{\ell-1}$};
  \foreach \i/\j in {1/1,2/2,4/3,5/4,6/5} {
    \node[ov/block,draw=ovTeal,fill=ovTeal!6,font=\footnotesize,
      minimum width=.83cm,minimum height=.55cm] (n\i) at (r\i) {$\sB_{i_\j}$};
  }
  \node[ov/block,draw=ovTeal,fill=ovTeal!6,
    minimum width=1.10cm,minimum height=.92cm] (n0) at (r0) {};
  \node[ov/block,draw=ovTeal,fill=ovTeal!6,
    minimum width=1.10cm,minimum height=.92cm] (n8) at (r8) {};
  \node[font=\footnotesize] at ($(r0)+(0,.23)$) {$\sB_0$};
  \node[font=\footnotesize] at ($(r8)+(0,.23)$) {$\sB_{\ell/2}$};
  \coordinate (x) at ($(r0)+(.13,-.16)$);
  \coordinate (y) at ($(r8)+(-.13,-.16)$);
  \draw[ov/query] (x)--(y);
  \node[ov/endpoint] at (x) {};
  \node[ov/endpoint] at (y) {};
  \node[font=\footnotesize,text=ovBlue] at ($(r0)+(-.24,-.16)$) {$x$};
  \node[font=\footnotesize,text=ovBlue] at ($(r8)+(.24,-.16)$) {$y$};
  \node[ov/label,text=ovBlue] at (0,.10) {$q=xy$};
  \foreach \i/\j in {0/1,1/2,4/5,5/6} {
    \draw[ov/route] (n\i) to[bend left=15] (n\j);
  }
  \draw[ov/route] (n2.north) .. controls (-2.45,3.05) and (-1.00,3.25) .. (n4.north west);
  \draw[ov/route] (n6.east) .. controls (3.85,1.95) and (4.30,1.30) .. (n8.north east);
  \foreach \labeltext/\xx/\yy in {1/-4.30/.80,2/-3.60/1.75,3/-1.70/3.36,
                                 4/.85/3.10,5/2.38/2.80,6/4.62/1.45} {
    \node[ov/label,font=\footnotesize,text=ovTeal] at (\xx,\yy) {$L_{\labeltext}$};
  }
  \node[ov/note] at (.20,-3.25) {green arrows: successful transitions \qquad grey blocks: not visited};
\end{tikzpicture}}
    \caption{\textbf{Routing along the half-cycle.} An illustrative route on the block cycle.
     Green arrows are successful transitions, and dark-orange lines indicate the corresponding cross-block edges.}\label{fig:overview-routing}
  \end{minipage}
  \par\addvspace{8pt}
}

\newcommand{\OverviewAnchorFigure}{%
  \par\addvspace{8pt}\noindent
  \begin{minipage}{\linewidth}
    \centering
    \captionsetup{type=figure,font=small,skip=4pt}
    \captionsetup[subfloat]{font=small,labelfont=rm,textfont=rm}
    \subfloat[The anchored graph.\label{fig:overview-anchor-structure}]{%
      \begin{minipage}[b]{.48\linewidth}
        \centering
        \begin{tikzpicture}[ov/base]
  \path[use as bounding box] (-3.90,-1.98) rectangle (3.90,1.82);
  \draw[ov/region] (-3.75,-1.50) rectangle (-.15,1.64);
  \draw[ov/region] (.15,-1.50) rectangle (3.75,1.64);
  \node at (-1.95,1.33) {anchor $\sC$};
  \node at (1.95,1.33) {ring $\sR$};
  \foreach \i in {0,...,7} {
    \coordinate (a\i) at ({-1.95+1.25*cos(90+45*\i)},{-.13+1.02*sin(90+45*\i)});
  }
  \foreach \i in {0,...,7} {
    \foreach \j in {0,...,7} {
      \ifnum\j>\i \draw[ov/local] (a\i)--(a\j);\fi
    }
  }
  \foreach \i in {0,...,7} {\node[ov/vertex] at (a\i) {};}
  \begin{scope}[shift={(1.95,-.12)},scale=1.10,shift={(-1.95,.12)},transform shape]
    \foreach \i in {0,...,7} {
      \coordinate (b\i) at ({1.95+1.04*cos(180-45*\i)},{-.12+.86*sin(180-45*\i)});
    }
    \foreach \i in {0,...,7} {
      \foreach \j in {0,...,7} {
        \ifnum\j>\i
          \pgfmathtruncatemacro{\dist}{min(\j-\i,8-\j+\i)}
          \ifnum\dist=1
            \draw[ovMuted!85,line width=.9pt] (b\i)--(b\j);
          \else
            \draw[ovMuted!40,line width=.3pt] (b\i)--(b\j);
          \fi
        \fi
      }
    }
    \foreach \i in {0,...,7} {
      \node[draw=ovMuted!70,fill=white,rounded corners=1pt,
        minimum width=.42cm,minimum height=.34cm,inner sep=0pt] at (b\i) {};
    }
    \draw[ov/query] (b0)--(b4);
    \node[ov/endpoint] at (b0) {};
    \node[ov/endpoint] at (b4) {};
    \node[font=\footnotesize,text=ovBlue,inner sep=1pt] at (1.95,.12) {$q$};
  \end{scope}
\end{tikzpicture}
      \end{minipage}%
    }\hfill
    \subfloat[Anchor/star representation.\label{fig:overview-anchor-decomposition}]{%
      \begin{minipage}[b]{.48\linewidth}
        \centering
        \begin{tikzpicture}[ov/base]
  \path[use as bounding box] (-3.90,-1.98) rectangle (3.90,1.82);
  \draw[ov/region] (-3.75,-1.50) rectangle (-.50,1.64);
  \draw[ov/region] (.50,-1.50) rectangle (3.75,1.64);
  \node at (-2.13,1.33) {anchor $\sC$};
  \node at (2.13,1.33) {ring $\sR$};
  \coordinate (u) at (1.22,.34);
  \coordinate (v) at (3.00,-.35);
  \foreach \i in {0,...,7} {
    \coordinate (a\i) at ({-2.13+1.18*cos(90+45*\i)},{-.12+1.00*sin(90+45*\i)});
  }
  \foreach \i in {0,...,7} {
    \draw[ov/star faint] (u)--(a\i) (v)--(a\i);
    \foreach \j in {0,...,7} {
      \ifnum\j>\i \draw[ov/local faint,draw=ovOrange!55,line width=.6pt] (a\i)--(a\j);\fi
    }
  }
  \draw[ov/local] (a0)--(a2) (a1)--(a5) (a3)--(a4) (a4)--(a6);
  \draw[ov/star] (u)--(a0) (u)--(a6) (v)--(a5);
  \draw[ov/query] (u)--(v);
  \node[text=ovBlue,anchor=south,inner sep=1pt] at (2.3,.05) {$\va_{uv}$};
  \node[ov/endpoint] at (u) {};
  \node[ov/endpoint] at (v) {};
  \node[font=\footnotesize,anchor=south,inner sep=.3pt] at ($(u)+(0,.20)$) {$u$};
  \node[font=\footnotesize,anchor=north,inner sep=.3pt] at ($(v)+(0,-.20)$) {$v$};
  \foreach \i in {0,...,7} {\node[ov/vertex] at (a\i) {};}
  \draw[ov/local] (-3.30,-1.80)--(-2.75,-1.80);
  \node[ov/key] at (-2.60,-1.80) {anchor columns};
  \draw[ov/star] (.45,-1.80)--(1.00,-1.80);
  \node[ov/key] at (1.15,-1.80) {star columns};
\end{tikzpicture}
      \end{minipage}%
    }
    \caption{\textbf{Adding the anchor.}
     (a) The anchor clique $\sC$ and the ring $\sR$ are disconnected and have the same number of vertices.
     (b) For an anchor-good ring edge $uv$, the long-dashed blue column $\va_{uv}$ is represented by orange anchor columns and short-dashed green star columns.}\label{fig:overview-anchor}
  \end{minipage}
  \par\addvspace{8pt}
}

\OverviewRingFigure

\paragraph{From statistical distance to leverage scores.}
Fix an incidence sketch matrix $\mB$ and the vertex permutation $\pi$.
The query increment shifts the sketch by its column $\vb_q$.
Intuitively, if this column can be represented by other columns with coefficients small relative to their edge weights, then small shifts in the existing noise can produce the same change in the sketch, making the query increment difficult to detect.
Specifically, by bounding the Hellinger distance under these shifts, we reduce the TV distance to the following leverage score of $\vb_q$ with respect to other columns of $\mB$ weighted by the base weights:
\begin{align*}
	\lev_{\vb_q}\Bigl(\bigl(\mB \diag(\vw)\bigr)_{-q}\Bigr) = \min\biggl\{\sum_{e \in \binom{\sR}{2} \setminus \{q\}}\frac{f_e^2}{w_e^2}: \vb_q = \sum_{e \in \binom{\sR}{2} \setminus \{q\}} f_e \vb_e\biggr\},
\end{align*}
where $\sR$ is the ring vertex set and $\vw = (w_e)$ is the base-weight vector.
Therefore, it suffices to make the expectation of this leverage score, truncated at one, small over $\pi$.
Here, the truncation allows rare failures of the representation.
Thus, by the above expression for the leverage score, our goal reduces to finding a decomposition of $\vb_q$ via other columns in $\mB$ that uses few columns between distant blocks, with controlled coefficients, since these columns have small base weights.

\paragraph{A first decomposition.}
We first seek to represent $\vb_{xy}$ using columns for edges within a separate reserve set $\sT_0 \subseteq \sB_0$, together with edges from $x$ and $y$ to this reserve, as illustrated in \cref{fig:overview-reserve}.
After this decomposition, the edges within $\sT_0$ and from $x$ to $\sT_0$ stay within block $0$, and only the star from $y$ to $\sT_0$ crosses to a distant block.

To obtain this representation, we use the following \emph{leverage-score averaging argument}.
The sum of the leverage scores of a matrix's columns equals its rank.
Consequently, if the number of columns is much larger than the rank, a uniformly chosen column can be represented by the others with small coefficient norm with high probability (see \cref{lem:large_leverage_bound}).
We apply this fact to the clique on $\sT_0 \cup \{x,y\}$.
The incidence property ensures that its columns are supported only on measurements assigned to these vertices.
Therefore, random labeling bounds its expected rank by the size of this vertex set times the average number of measurements per vertex, whereas its number of columns is quadratic in the set size.
Hence, taking a sufficiently large reserve gives, with high probability, the desired representation with small coefficient norm.

More generally, we designate a reserve $\sT_i$ in each block and apply the same argument to a vertex $v$ and a reserve $\sT_i$ with $v,y \notin \sT_i$.
Whenever the above representation succeeds, we write
\begin{align*}
	\vb_{vy} = \mB_{v, \sT_i} \valpha^{(v)}_{v, \sT_i} + \mB_{y, \sT_i} \valpha^{(v)}_{y, \sT_i} + \mB_{\sT_i} \valpha^{(v)}_{\sT_i},
\end{align*}
where the three matrices contain columns from $v$ to the reserve, from $y$ to the reserve, and within the reserve, respectively.
We call the middle term the \emph{root-star term}.

\paragraph{Moving the root-star term.}
We use telescoping to cancel intermediate root-star terms, moving from the star between $y$ and $\sT_0$ to one within the block containing $y$.
Consider first the transition from $\sT_0$ to $\sT_1$ in \cref{fig:overview-transition}.
We introduce an auxiliary set $\sS_0 \subseteq \sB_0$, disjoint from $\sT_0 \cup \{x\}$.
Using the same leverage-score averaging argument, we represent the difference between the two root-star terms for $x$ by the corresponding differences for vertices $v \in \sS_0$, with small coefficient norm.
For each such $v$, subtracting its two decompositions of $\vb_{vy}$ cancels this column.
Thus, the difference between the purple root-star terms is represented by the orange within-block columns and the dark-orange cross-block columns between $\sS_0$ and $\sT_1$.
This gives a transition with small cross-block coefficient norm and no representing column incident to $y$.

\OverviewTransitionFigure

This construction extends to transitions from $\sT_i$ to $\sT_j$ using auxiliary vertices in $\sS_i$.
Successive transitions telescope, as illustrated in \cref{fig:overview-second-transition}, until the remaining root-star term lies within the block of $y$.
Together with the initial within-block terms, they give a representation of $\vb_{xy}$ whose cross-block cost is determined by the chosen transitions.

\OverviewSecondTransitionFigure

\paragraph{Routing on the block cycle.}
Viewing successful transitions as directed edges between blocks, our goal is to find an increasing path from block $0$ to block $\ell / 2$ with small total distance-weighted cost; see \cref{fig:overview-routing}.
Most blocks have few missing incoming and outgoing transitions, so intuitively the path should rarely need long jumps as the blocks are permuted randomly.
We prove that a path visiting as many of these blocks as possible has the required guarantee: the expected sum of the eighth powers of its step lengths is $O(\ell)$, matching the cost induced by our edge-weight scaling.
This gives the small leverage-score bound needed for the $\Omega\bigl(n^{6/5}\bigr)$ lower bound for incidence sketches.

\OverviewRoutingFigure

\paragraph{From arbitrary sketches to incidence sketches.}
To handle an arbitrary sketch matrix $\mA$ of dimension $N$, we add a randomly labeled anchor clique $\sC$ with heavy edge weights, disconnected from the ring; see \cref{fig:overview-anchor-structure}.
This preserves the spectral distinction between the hard distributions.

Write $\mA_{\sC}$ for the anchor columns.
Our goal is to decompose most ring columns as
\begin{align*}
	\va_{uv} = \mT \vb_{uv} + \mA_{\sC} \vc_{uv},
\end{align*}
where $\mT$ is a common transformation, $\vb_{uv}$ is supported only on coordinates assigned to its endpoints $u,v$, and $\vc_{uv}$ has small norm.
The columns $\vb_{uv}$ then form an incidence sketch $\mB$.

Leverage-score averaging first represents most ring columns using the anchor columns and the two endpoint stars $\mA_{u, \sC}$ and $\mA_{v, \sC}$, with small coefficient norm; see \cref{fig:overview-anchor-decomposition}.
These stars are used only in the analysis, since their edges are absent from the input graph.

\OverviewAnchorFigure

To construct $\mT$, for each ring vertex $v$ let $\mT_v$ contain orthonormal eigenvectors corresponding to the positive eigenvalues of $\mA_{v, \sC} \mA_{v, \sC}^{\top} - \mA_{\sC} \mA_{\sC}^{\top}$, scaled by the square roots of their eigenvalues.
We concatenate them as $\mT \coloneqq \bigl[\mT_v: v \in \sR\bigr]$, assigning the coordinates associated with $\mT_v$ to vertex $v$.
The leverage-score characterization in \cref{lem:leverage-Loewner} gives the desired decomposition, with $\vb_{uv}$ supported on the coordinate blocks for $u$ and $v$.
Using Sylvester's law of inertia and a trace--leverage bound, we prove that the expected number of columns of $\mT$, and hence the expected dimension of $\mB$, is $O(N)$.

\paragraph{Lifting the representation.}
Since $\mT$ is common to all columns, a representation of $\vb_q$ with small coefficients lifts to a representation of $\va_q$ with the same ring coefficients, plus an anchor residual.
The coefficient bounds and heavy anchor weights make the residual's weighted cost negligible.
Thus the incidence argument yields the $\Omega\bigl(n^{6/5}\bigr)$ lower bound for arbitrary sketches with polynomially bounded edge weights.

\paragraph{Comparison with the lower bound proof in \cite{CKL22weighted}.}

Chen, Khanna, and Li~\cite{CKL22weighted} use randomly labeled block-cycle graphs with Gaussian weights and an additional crossing edge.
Crucially, in their block-cycle graphs, there are no edges between nonadjacent blocks.
In their proof, they reduce incidence sketches to a more structured class of sketches called signed sketches, for which the relevant leverage score becomes a matrix-weighted effective resistance.
They then bound this resistance using matrix-weighted extensions of expander decomposition and the preservation of expansion under vertex sampling.

Instead of reducing incidence sketches to signed sketches, we exploit the incidence structure more directly to bound the general leverage score using our leverage-score averaging and routing arguments.
Also, from the perspective of our routing argument, restricting transitions to adjacent blocks would leave little flexibility when a transition fails.
We therefore complete the block cycle to a clique, allowing transitions between nonadjacent blocks and routing around unsuccessful transitions.
Intuitively, this makes it easier to construct a successful routing decomposition, and our analysis shows that the cost incurred by using longer transitions can still be controlled.

We also use shifted Beta noise to keep every realization's weights within a positive polynomial range.
The smooth decay of the Beta noise at the boundary also gives the quadratic Hellinger-distance bound under shifts that allows us to control TV distance by weighted leverage scores.
Finally, the anchor decomposition extends our lower bound result to arbitrary sketches.

\paragraph{Paper organization.}
\Cref{sec:prelim} collects the facts about leverage scores and statistical distance used in the proof.
\Cref{sec:hard-distributions} constructs the hard distributions and reduces indistinguishability to a leverage-score bound.
\Cref{sec:incidence_sketch} proves this bound for incidence sketches through the decomposition and routing argument.
Finally, \cref{sec:arbitrary-sketch} develops the anchor reduction and lifting argument to prove the lower bound for arbitrary linear sketches.

\section{Preliminaries} \label{sec:prelim}

\paragraph{Basic notation.}
For any positive integer $n$, we write $[n] \coloneqq \{1,2,\dots,n\}$.
For a vector $\vx \in \R^m$, we write $\diag(\vx)$ for the $m \times m$ diagonal matrix whose $i$-th diagonal entry is $x_i$.
For a matrix $\mM$, we use $\|\mM\|_2$ to denote its spectral norm and $\mM^{\dagger}$ to denote its Moore--Penrose pseudoinverse.
For a matrix $\mM$ with $m$ columns and $i \in [m]$, we write $\mM_{-i}$ for the matrix obtained by removing the $i$-th column of $\mM$.
We write $\binom{S}{2}$ for the set of unordered pairs of distinct elements in a set $S$.
For two vertices $u,v$ on the graph, we write $uv$ for the undirected edge slot between $u$ and $v$.

\paragraph{Loewner order.}
For real symmetric matrices $\mM,\mN$ of the same dimension, we write $\mM \preceq \mN$ if $\mN - \mM$ is positive semidefinite (PSD), i.e., if $\vx^{\top} (\mN - \mM) \vx \ge 0$ for every vector $\vx$.
For any matrix $\mU$ of compatible dimensions, $\mM \preceq \mN$ implies $\mU \mM \mU^\top \preceq \mU \mN \mU^\top$.

\subsection{Leverage Scores}

For a target vector $\vv$ and a collection of vectors $(\vu_i)_{i = 1}^{m}$ of the same dimension as $\vv$, we define the \textit{leverage score} of $\vv$ with respect to $(\vu_i)_{i = 1}^{m}$ as
\begin{align} \label{eqn:target-lev}
	\lev_{\vv}\bigl((\vu_i)_{i = 1}^{m}\bigr) \coloneqq \min\biggl\{\sum_{i = 1}^{m} c_i^2: \vv = \sum_{i = 1}^{m} c_i \vu_i\biggr\},
\end{align}
which is interpreted as $+\infty$ if $\vv \notin \range\bigl([\vu_1 \  \cdots \  \vu_m]\bigr)$.
For a matrix $\mU$, we write $\lev_{\vv}(\mU)$ for the leverage score of $\vv$ with respect to the columns of $\mU$.
For $\mU = [\vu_1 \  \cdots \  \vu_m]$, we have $\sum_{i=1}^{m} \lev_{\vu_i}(\mU) = \rank(\mU)$, and also $\lev_{\vv}(\mU) = \vv^{\top} (\mU \mU^{\top})^{\dagger} \vv$ for any $\vv \in \range(\mU)$.

\begin{lemma} \label[lemma]{lem:large_leverage_bound}
	Let $\mU = [\vu_1 \  \cdots \  \vu_m]$.
	If $I$ is uniformly randomly chosen among $[m]$, then
	\begin{align*}
		\Pr\bigl\{\lev_{\vu_I}(\mU_{-I}) > 1\bigr\} \le \frac{2\rank(\mU)}{m}.
	\end{align*}
\end{lemma}

\begin{proof}
	We first relate $\lev_{\vu_i}(\mU_{-i})$ to $\lev_{\vu_i}(\mU)$, for $i \in [m]$.
	Suppose that $\lev_{\vu_i}(\mU_{-i})$ is finite.
	If the coefficient of $\vu_i$ in a representation of itself is fixed to $a$,
	the remaining columns must represent $(1-a)\vu_i$.
	By homogeneity of the minimum squared coefficient norm,
	\begin{align*}
		\lev_{\vu_i}(\mU) & = \min_{a \in \R} \bigl\{a^2 + (1-a)^2 \lev_{\vu_i}(\mU_{-i})\bigr\} = \frac{\lev_{\vu_i}(\mU_{-i})}{1+\lev_{\vu_i}(\mU_{-i})},
	\end{align*}
	where the right-hand side is interpreted as $1$ when the leave-one-out value is infinite.
	If the leave-one-out value is infinite, every feasible representation
	must give $\vu_i$ coefficient $1$; otherwise, it would yield a representation of $\vu_i$ using only the remaining columns.
	Taking all other coefficients to be zero then shows that the ordinary leverage score equals $1$, as claimed.
	
	Thus, $\lev_{\vu_i}(\mU_{-i}) > 1$ is equivalent to $\lev_{\vu_i}(\mU) > 1/2$.
	Combining this with $\sum_{i=1}^{m} \lev_{\vu_i}(\mU) = \rank(\mU)$ and the uniformity of $I$ proves the lemma.
\end{proof}

\begin{lemma} \label[lemma]{lem:leverage-Loewner}
	$\lev_{\vv}(\mU) \le 1$ is equivalent to $\vv \vv^{\top} \preceq \mU \mU^{\top}$.
\end{lemma}

\begin{proof}
	First suppose that $\lev_{\vv}(\mU) \le 1$.
	Then there exists a coefficient vector $\vc$ with $\|\vc\|_2 \le 1$ such that $\vv = \mU \vc$.
	Thus, $\vc \vc^{\top} \preceq \mI$ and $\vv \vv^{\top} = \mU \vc \vc^{\top} \mU^{\top} \preceq \mU \mU^{\top}$.
	On the other hand, suppose that $\vv \vv^{\top} \preceq \mU \mU^{\top}$.
	This implies that
	\begin{align*}
		\bigl(\vv^{\top} (\mU \mU^{\top})^{\dagger} \vv\bigr)^2 = \vv^{\top} (\mU \mU^{\top})^{\dagger} \vv \vv^{\top} (\mU \mU^{\top})^{\dagger} \vv \le \vv^{\top} (\mU \mU^{\top})^{\dagger} \mU \mU^{\top} (\mU \mU^{\top})^{\dagger} \vv = \vv^{\top} (\mU \mU^{\top})^{\dagger} \vv,
	\end{align*}
	which gives $\vv^{\top} (\mU \mU^{\top})^{\dagger} \vv \le 1$.
	The relation $\vv \vv^{\top} \preceq \mU \mU^{\top}$ also implies that $\vv \in \range(\mU)$, so $\lev_{\vv}(\mU) = \vv^{\top} (\mU \mU^{\top})^{\dagger} \vv \le 1$, completing the proof.
\end{proof}

\subsection{Total Variation and Hellinger Distances}

For probability distributions $P, Q$ on the same measurable space, their \textit{total variation distance} (\textit{TV distance}) is
\begin{align*}
	\TV(P, Q) \coloneqq \sup_A|P(A) - Q(A)|.
\end{align*}
If a sample is drawn from $P$ with probability $1/2$ and from $Q$ otherwise, then any randomized algorithm for determining which distribution generated the sample satisfies
\begin{equation} \label{eqn:TV-distinguishing}
	\Pr[\text{correct identification}] \le \frac{1}{2} + \frac{1}{2} \TV(P,Q).
\end{equation}
For random variables $X,Y$, we write $\TV(X,Y)$ for the TV distance between their distributions, and write $\TV(X,Y \mid U)$ for the distance between their conditional distributions given the realized value of $U$.

Let $p,q$ be densities of $P,Q$ with respect to a common measure $\mu$, meaning that $P(A) = \int_A p \, \mathrm{d}\mu$ and $Q(A) = \int_A q \, \mathrm{d}\mu$ for every measurable set $A$.
The square of the \emph{Hellinger distance} between $P$ and $Q$ is defined by
\begin{align*}
	H^2(P,Q) \coloneqq \frac{1}{2} \int \bigl(\sqrt{p} - \sqrt{q}\bigr)^2 \, \mathrm{d}\mu.
\end{align*}
This definition is independent of the choice of $\mu$, and one may always take $\mu = P+Q$.
For random variables $X$ and $Y$, we use $H(X,Y)$ to denote the Hellinger distance between their distributions.
For probability distributions $P_1,\dots,P_m$, the notation $\bigotimes_{i=1}^{m} P_i$ denotes their product distribution, i.e., the joint distribution of independent random variables $X_i \sim P_i$.

We will use the following standard properties of the Hellinger distance.

\begin{lemma} \label[lemma]{lem:hellinger-products}
	For probability distributions $P_i,Q_i$ on the same measurable space for each $i \in [m]$,
	\begin{align*}
		H^2\biggl(\bigotimes_{i = 1}^{m} P_i, \bigotimes_{i = 1}^{m} Q_i\biggr) \le \sum_{i = 1}^{m} H^2(P_i,Q_i).
	\end{align*}
	For probability distributions $P,Q$ on the same measurable space, $\TV(P,Q) \le \sqrt{2} \, H(P,Q)$.
\end{lemma}

\section{Hard Distributions and a Reduction to Leverage Scores} \label{sec:hard-distributions}

\subsection{The Hard Distributions}

In this subsection, we describe the hard distributions that we will use to prove the lower bound for incidence sketches.

Given an integer $b \ge 1$ and an even integer $\ell \ge 2$, we set $n \coloneqq \ell b$ and let the graph consist of the vertex set $\sR \coloneqq [n]$.
We start from a fixed template of $\ell$ cyclically ordered blocks, indexed by $0,\dots,\ell-1$, so that block $i$ contains $b$ vertex slots $ib+1,\dots,(i+1)b$.
We also arbitrarily fix one distinguished query slot in block $0$ and another in block $\ell / 2$.

Let $\pi$ be a uniformly random permutation of $[n]$, and place vertex $\pi(t)$ in slot $t$ for each $t \in [n]$.
Thus, the resulting vertex blocks are
\begin{align*}
	\sB_i \coloneqq \bigl\{\pi(ib+j): j \in [b]\bigr\}, \qquad i=0,\dots,\ell-1.
\end{align*}
Let $x$ and $y$ be the vertices occupying the distinguished query slots in blocks $0$ and $\ell/2$, respectively, and define the query edge $q \coloneqq xy$.
Thus, both the vertex partition and the query edge are determined by $\pi$.

For block indices $0 \le i,j < \ell$, let $L(i,j) \coloneqq \min\bigl\{|i-j|, \ell-|i-j|\bigr\}$ denote the cyclic block distance.
We define the \emph{base weight} of each edge by
\begin{align} \label{eqn:physical-scales}
	w_e \coloneqq
	\begin{cases}
		w_{\mathrm{int}}, & e \in \binom{\sB_i}{2} \text{ for some }i, \\
		w_{\mathrm{ext}} \, L(i,j)^{-4}, & e = uv, u \in \sB_i, v \in \sB_j, i \ne j,
	\end{cases}
\end{align}
where $w_{\mathrm{int}} \coloneqq n^2$ and $w_{\mathrm{ext}} \coloneqq \ell / \bigl(16 b^2\bigr)$.
We write $\vw \coloneqq (w_e)_{e \in \binom{\sR}{2}}$ for the base-weight vector, which is deterministic conditional on $\pi$.

We set $Z$ as $1/2$ plus a random variable that follows the $\operatorname{Beta}(3,3)$ distribution.
Equivalently, $Z$ has density
\begin{equation} \label{eqn:beta-density}
	f(z) \coloneqq \mathbf{1}\Bigl\{\frac{1}{2} \le z \le \frac{3}{2}\Bigr\} \cdot 30 \Bigl(z - \frac{1}{2}\Bigr)^2 \Bigl(\frac{3}{2} - z\Bigr)^2.
\end{equation}
It holds that $Z \in [1/2, 3/2]$ and $\E[Z] = 1$.
Then, conditional on $\pi$, we independently sample copies $Z_e$ of $Z$ for each $e \in \binom{\sR}{2}$.
Independently of $\pi$ and the $Z_e$'s, we further randomly choose the $\Yes$ or $\No$ case, each with probability $1/2$.
The \textit{perturbed weights} of the input graph are then set as
\begin{align*}
	\tw_e \coloneqq
	\begin{cases}
		w_e Z_e, & \text{in the } \No \text{ case}, \\
		w_e Z_e + \mathbf{1}\{e = q\}, & \text{in the } \Yes \text{ case}.
	\end{cases}
\end{align*}
The input graph is the complete graph on $\sR$ with weight vector $\tvw \coloneqq (\tw_e)_{e \in \binom{\sR}{2}}$.

By our setting of the edge weights, we have the following bound on the positive-weight ratio of the input graph.

\begin{lemma} \label[lemma]{lem:weight-range}
	Every graph sampled under $\Yes$ or $\No$ has positive-weight ratio $O\bigl( n^5 \bigr)$.
\end{lemma}

\begin{proof}
	For an edge $e$ joining distinct blocks $\sB_i$ and $\sB_j$, the cyclic block distance satisfies $1 \le L(i,j) \le \ell/2$.
	Thus, its base weight satisfies
	\begin{align*}
		w_e = \frac{\ell}{16 b^2} \, L(i,j)^{-4} \ge \frac{\ell}{16 b^2} \left(\frac{\ell}{2}\right)^{-4} = \frac{1}{\ell^3 b^2} \ge \frac{1}{n^3}.
	\end{align*}
	This lower bound also holds for edges within a block, whose base weight is $n^2$.
	Moreover, every cross-block base weight is at most $\ell / \bigl(16 b^2\bigr) \le n^2$.
	Hence every edge satisfies $1 / n^3 \le w_e \le n^2$.

	In either the $\Yes$ or the $\No$ case, each base weight is multiplied by a noise in $[1/2,3/2]$, and the query-edge weight may additionally be increased by one.
	Consequently,
	\begin{align*}
		\frac{1}{2 n^3} \le \frac{1}{2} w_e \le \tw_e \le \frac{3}{2} w_e + 1 \le 2 n^2.
	\end{align*}
	Therefore, every realization of either distribution has positive-weight ratio at most $2 n^2 / \bigl(1 / \bigl(2 n^3\bigr)\bigr) = 4 n^5$, which proves the lemma.
\end{proof}

The next lemma shows that a $(1 \pm 0.01)$-spectral sparsifier of the graph, together with $\pi$, can help to distinguish the $\Yes$ and $\No$ distributions.

\begin{lemma} \label[lemma]{lem:spectral-separation}
	Suppose a linear sketch computes a $(1 \pm 0.01)$-spectral sparsifier with probability at least $0.99$ on every input in the support of $\No$ or $\Yes$.
	Then, given the output $\widehat{\mL}$ of the recovery algorithm of the linear sketch and the permutation $\pi$, one can identify whether the input was sampled from $\Yes$ or $\No$ with success probability at least $0.99$ under either distribution.
\end{lemma}

\begin{proof}
	We construct a test potential vector $\vphi \in \R^n$, calculate its quadratic form $\vphi^{\top} \mL_{\tvw} \vphi$ under the two input distributions, and then show that the spectral approximation guarantee of the linear sketch allows us to distinguish between the two cases.

	Let $\phi_i \coloneqq 2 \min\{i, \ell - i\} / \ell$, so $|\phi_i - \phi_j| \le 2 L(i,j) / \ell$.
	For each $0 \le i < \ell$, set $\phi(v) \coloneqq \phi_i$ for all $v \in \sB_i$, and write $\vphi$ for the resulting vector.
	First consider the $\No$ case.
	For $u \in \sB_i$ and $v \in \sB_j$ with $i \ne j$, the perturbed weight satisfies $\tw_{uv} = w_{uv} Z_{uv} = Z_{uv} \cdot \ell / \bigl(16 b^2\bigr) \cdot \, L(i,j)^{-4} \le (3/2) \cdot \ell / \bigl(16 b^2\bigr) \cdot \, L(i,j)^{-4}$.
	Edges within a block contribute zero to the quadratic form, so
	\begin{align*}
		0 \le \vphi^{\top} \mL_{\tvw} \vphi & \le \frac{3 \ell}{32 b^2} \sum_{0 \le i < j < \ell} \sum_{u \in \sB_i} \sum_{v \in \sB_j} L(i, j)^{-4} \, (\phi_i - \phi_j)^2 \\
		& \le \frac{3 \ell}{32 b^2} \sum_{0 \le i < j < \ell} b^2 L(i, j)^{-4} \, \frac{4 L(i,j)^2}{\ell^2} \\
		& = \frac{3}{8 \ell} \sum_{0 \le i < j < \ell} L(i, j)^{-2} \le \frac{3}{8 \ell} \sum_{d \ge 1} \ell \cdot \frac{1}{d^2} \le \frac{3}{4}.
	\end{align*}
	In the penultimate step, we used the fact that, for each $d \ge 1$, there are at most $\ell$ pairs $0 \le i < j < \ell$ with $L(i,j) = d$.
	Under $\Yes$, the extra unit of weight on the query edge makes $\vphi^{\top} \mL_{\tvw} \vphi$ at least $1$, since $\phi(x) = 0$ and $\phi(y) = 1$.

	Now we consider the following approach to distinguish the two distributions.
	Given $\pi$ and the output $\widehat{\mL}$ of the recovery algorithm, compute $\vphi^{\top} \widehat{\mL} \vphi$.
	Declare $\Yes$ if $\vphi^{\top} \widehat{\mL} \vphi > 7/8$ and $\No$ otherwise.

	If the recovery algorithm outputs a valid $(1 \pm 0.01)$-spectral approximation $\widehat{\mL}$ to the input Laplacian $\mL_{\tvw}$, then $\vphi^\top \widehat{\mL} \vphi \le 1.01 \, (3/4) < 7/8$ in the $\No$ case, whereas $\vphi^\top \widehat{\mL} \vphi \ge 0.99 > 7/8$ in the $\Yes$ case.
	By assumption, recovery succeeds with probability at least $0.99$ on every input in the support of $\No$ or $\Yes$, over the sketch matrix and the recovery algorithm's randomness.
	Since the distinguishing algorithm is correct whenever recovery succeeds, averaging over inputs gives success probability at least $0.99$ under either distribution.
\end{proof}

\subsection{Reduction to Leverage Scores}

To upper bound the TV distance, we will use the following elementary bound on the Hellinger distance between a distribution and its shifted version.

\begin{lemma} \label[lemma]{lem:hellinger-shift}
	Let $X$ be a real-valued random variable with density $p$ on $\R$, such that $u \coloneqq \sqrt{p}$ is absolutely continuous and $\int_{\R}|u'(z)|^2 \, \mathrm{d}z < \infty$.
	For every $h \in \R$,
	\begin{align*}
		H^2(X,X+h) \le \frac{1}{2} h^2 \int_{\R} \bigl|u'(z)\bigr|^2 \, \mathrm{d}z.
	\end{align*}
\end{lemma}

\begin{proof}
	The density of $X+h$ is $p(z-h)$, so the definition of Hellinger distance gives
	\begin{align*}
		2 H^2(X,X+h) = \int_{\R} \bigl|u(z) - u(z-h)\bigr|^2 \, \mathrm{d}z.
	\end{align*}
	By absolute continuity of $u$ and the fundamental theorem of calculus,
	\begin{align*}
		u(z) - u(z-h) = h \int_0^1 u'(z - t h) \, \mathrm{d}t.
	\end{align*}
	Substituting this identity and applying the Cauchy--Schwarz inequality to the integral over $t \in [0,1]$, we obtain
	\begin{align*}
		2 H^2(X,X+h) & = h^2 \int_{\R} \biggl|\int_{0}^{1} u'(z-th) \, \mathrm{d}t\biggr|^2 \, \mathrm{d}z \\
		& \le h^2 \int_{\R} \int_{0}^{1} \bigl|u'(z - t h)\bigr|^2 \, \mathrm{d}t \, \mathrm{d}z = h^2 \int_{0}^{1} \int_{\R} \bigl|u'(z - t h)\bigr|^2 \, \mathrm{d}z \, \mathrm{d}t \\
		& = h^2 \int_{0}^{1} \biggl(\int_{\R} \bigl|u'(z)\bigr|^2 \, \mathrm{d}z\biggr) \, \mathrm{d}t = h^2 \int_{\R} \bigl|u'(z)\bigr|^2 \, \mathrm{d}z,
	\end{align*}
	where the order of integration is exchanged in the second line by Tonelli's theorem, since the integrand is nonnegative.
	This completes the proof.
\end{proof}

The following lemma upper bounds the TV distance between the outputs of the linear sketch under the $\Yes$ and $\No$ distributions by a leverage score.

For a statistic $X$ of this experiment, write $(X)_{\Yes}$ and $(X)_{\No}$ for its distributions under the two cases, and $(X)_{\pi,\Yes}$ and $(X)_{\pi,\No}$ when additionally conditioning on the realized permutation.

\begin{lemma} \label[lemma]{lem:TV-leverage}
	For any fixed sketch matrix $\mA$ and realization of the random permutation $\pi$,
	\begin{align*}
		\TV\bigl((\mA \tvw)_{\pi, \Yes}, (\mA \tvw)_{\pi, \No}\bigr) \le \sqrt{\min\Bigl\{1, 10 \lev_{\va_q}\Bigl(\bigl(\mA \diag(\vw)\bigr)_{-q}\Bigr) \Bigr\}}.
	\end{align*}
\end{lemma}

\begin{proof}
	Note that for fixed $\pi$, $q$ and $\vw$ are deterministic.
	If $\va_q$ is not in the range of $\mA_{-q}$, the claim holds trivially since the TV distance is at most $1$.
	Thus, assume that $\va_q$ is in the range of $\mA_{-q}$.
	Recall that
	\begin{align*}
		\lev_{\va_q}\Bigl(\bigl(\mA \diag(\vw)\bigr)_{-q}\Bigr) & = \min\biggl\{ \sum_{e \in \binom{\sR}{2} \setminus \{q\}} c_e^2: \va_q = \sum_{e \in \binom{\sR}{2} \setminus \{q\}} c_e w_e \va_e \biggr\} \\
		& = \min\biggl\{ \sum_{e \in \binom{\sR}{2} \setminus \{q\}} \frac{c_e^2}{w_e^2}: \va_q = \sum_{e \in \binom{\sR}{2} \setminus \{q\}} c_e \va_e \biggr\}.
	\end{align*}
	Hence, there exist coefficients $(c_e)_{e \in \binom{\sR}{2} \setminus \{q\}}$ such that $\va_q = \sum_{e \in \binom{\sR}{2} \setminus \{q\}} c_e \va_e$ and $\sum_{e \in \binom{\sR}{2} \setminus \{q\}} c_e^2 / w_e^2 = \lev_{\va_q}\bigl(\bigl(\mA \diag(\vw)\bigr)_{-q}\bigr)$.
	Let $c_q \coloneqq 0$ and $\vc \coloneqq (c_e)_{e \in \binom{\sR}{2}}$, so $\mA \vc = \va_q$.

	Let $\tvw' = (\tw'_e)_{e \in \binom{\sR}{2}}$ have distribution $(\tvw)_{\pi,\No}$.
	Its coordinates are independent, with $\tw'_e = w_e Z_e$.
	We have
	\begin{align*}
		& \TV\bigl((\mA \tvw)_{\pi,\Yes}, (\mA \tvw)_{\pi,\No}\bigr) \\
		= {} & \TV(\mA \tvw' + \va_q, \mA \tvw') = \TV\bigl(\mA (\tvw' + \vc), \mA \tvw'\bigr) \\
		\le {} & \TV(\tvw', \tvw' + \vc) \le \sqrt{2} \, H(\tvw', \tvw' + \vc).
	\end{align*}
	Applying \cref{lem:hellinger-products} to all edges gives $H^2(\tvw', \tvw' + \vc) \le \sum_{e \in \binom{\sR}{2}} H^2(\tw'_e, \tw'_e + c_e)$.
	We next calculate each term $H^2(w_e Z_e, w_e Z_e + c_e)$ by \cref{lem:hellinger-shift}.
	The square root of the density of $Z$ is $u(z) \coloneqq \mathbf{1}\bigl\{\frac{1}{2} \le z \le \frac{3}{2}\bigr\} \sqrt{30} \bigl(z - \frac{1}{2}\bigr) \bigl(\frac{3}{2} - z\bigr)$, which is absolutely continuous.
	We have
	\begin{align*}
		\int_{\R} \bigl|u'(z)\bigr|^2 \, \mathrm{d}z = \int_{1/2}^{3/2} 30 (2-2z)^2 \, \mathrm{d}z = 10.
	\end{align*}
	Now, by \cref{lem:hellinger-shift}, for every base weight $w > 0$ and $c \in \R$ we have
	\begin{align*}
		H^2(wZ,wZ+c) = H^2\Bigl(Z,Z+\frac{c}{w}\Bigr) \le \frac{1}{2} \cdot \frac{c^2}{w^2} \int_{\R} \bigl|u'(z)\bigr|^2 \, \mathrm{d}z = \frac{5 c^2}{w^2}.
	\end{align*}
	Combining, we have
	\begin{align*}
		\TV\bigl((\mA\tvw)_{\pi,\Yes}, (\mA\tvw)_{\pi,\No}\bigr) \le \sqrt{10 \sum_{e \in \binom{\sR}{2} \setminus \{q\}} \frac{c_e^2}{w_e^2}} = \sqrt{10 \, \lev_{\va_q}\Bigl(\bigl(\mA \diag(\vw)\bigr)_{-q}\Bigr)},
	\end{align*}
	which together with the bound $1$ on the TV distance finishes the proof.
\end{proof}

The next lemma reduces the proof of our sketching dimension lower bounds to bounding the expected leverage score of the query column, truncated at one.
We additionally reveal $\pi$ so that the distinguishing algorithm can apply \cref{lem:spectral-separation} to identify the correct case whenever recovery succeeds.
For a fixed sketch matrix $\mA$, write $(\pi, \mA \tvw)_{\No}$ and $(\pi, \mA \tvw)_{\Yes}$ for the joint distributions of the permutation and the sketch under the two cases.

\begin{lemma} \label[lemma]{lem:reduction-to-leverage}
	Consider a linear sketch and suppose that, for every realization $\mA = \bigl[\va_e: e \in \binom{\sR}{2}\bigr]$ of its sketch matrix,
	\begin{align*}
		\E_{\pi}\Bigl[ \min\Bigl\{ 1, \lev_{\va_q}\Bigl(\bigl(\mA \diag(\vw)\bigr)_{-q}\Bigr) \Bigr\} \Bigr] < \frac{1}{100}.
	\end{align*}
	Then no recovery algorithm for this sketch can output a $(1 \pm 0.01)$-spectral sparsifier with probability at least $0.99$ on every input in the support of $\No$ or $\Yes$.
\end{lemma}

\begin{proof}
	Under the stated leverage score assumption, we first upper bound the success probability of any algorithm that distinguishes $\No$ from $\Yes$ given the sketch and $\pi$.
	We then show that a recovery algorithm satisfying the stated guarantee would contradict this bound.
	Fix a realization $\mA$ of the sketch matrix.
	Conditional on $\pi$, the sketch $\mA \tvw$ has distributions $(\mA \tvw)_{\pi,\No}$ and $(\mA \tvw)_{\pi,\Yes}$ under the two cases, respectively.
	Since $\pi$ has the same distribution under $\Yes$ and $\No$, the TV distance between the joint distributions equals the average conditional TV distance.
	By \cref{lem:TV-leverage},
	\begin{align*}
		& \TV\bigl((\pi,\mA \tvw)_{\Yes}, (\pi,\mA \tvw)_{\No}\bigr) \\
		= {} & \E_{\pi}\Bigl[\TV\bigl((\mA \tvw)_{\pi,\Yes}, (\mA \tvw)_{\pi,\No}\bigr)\Bigr] \\
		\le {} & \E_{\pi}\Biggl[\sqrt{\min\Bigl\{ 1, 10 \lev_{\va_q}\Bigl(\bigl(\mA \diag(\vw)\bigr)_{-q}\Bigr) \Bigr\}}\Biggr] \\
		\le {} & \sqrt{10 \E_{\pi}\Bigl[\min\Bigl\{ 1, \lev_{\va_q}\Bigl(\bigl(\mA \diag(\vw)\bigr)_{-q}\Bigr) \Bigr\}\Bigr]} < \frac{1}{\sqrt{10}}.
	\end{align*}
	The second inequality uses $\min\{1, 10 u\} \le 10\min\{1,u\}$ for $u \in [0,\infty]$ and Jensen's inequality.

	Since the two cases are equally likely, \eqref{eqn:TV-distinguishing} implies that any randomized distinguishing algorithm given the sketch and $\pi$ succeeds with probability at most $\frac{1}{2} (1 + 1 / \sqrt{10}) < \frac{3}{4}$.
	This bound holds for every fixed $\mA$.
	Since the sketch matrix is sampled independently of the input experiment, the same bound holds after averaging over its choice.

	Suppose now that a recovery algorithm satisfies the stated spectral guarantee on every input in the support of either distribution.
	Run it on $\mA$ and $\mA \tvw$, then apply the distinguishing algorithm from the proof of \cref{lem:spectral-separation} using $\pi$.
	The resulting algorithm is correct whenever recovery produces a valid spectral approximation.
	For each input, recovery succeeds with probability at least $0.99$ over the choice of sketch matrix and the recovery algorithm's randomness.
	Averaging over the input distribution therefore gives success probability at least $0.99$ in each case, contradicting the upper bound above.
\end{proof}

\section{Lower Bound for Incidence Sketches} \label{sec:incidence_sketch}

\begin{theorem} \label{thm:incidence_sketch}
	Any \emph{incidence sketch} of dimension $N$ that computes a $(1 \pm 0.01)$-spectral sparsifier with probability at least $0.99$ on every $n$-vertex graph with $w_{\max} / w_{\min} = O\bigl(n^5\bigr)$ must satisfy $N = \Omega\bigl(n^{6/5}\bigr)$.
\end{theorem}

The proof relies on the following representation of the query column, which we will also use in our proof for general linear sketches in \cref{sec:arbitrary-sketch}.

\begin{lemma} \label[lemma]{lem:endpoint-representation}
	Let $t$ be a positive integer and $\ell \ge 10$ be an even integer, and set $b \coloneqq \bigl(10^8 + 1\bigr)t+1$.
	Fix a vertex set $\sR$ of size $\ell b$ and an incidence sketch matrix $\mB = \bigl[\vb_e: e \in \binom{\sR}{2}\bigr]$ of dimension $N \le 10^{-9} \ell b t$.
	Under the uniform random labeling of the ring slots, there is an event $\event$ with $\Pr\{\event\} \ge 0.999$ on which the query column admits a representation
	\begin{align*}
		\vb_q = \sum_{e \in \binom{\sR}{2} \setminus \{q\}} f_e \vb_e.
	\end{align*}
	The coefficient vector $\vf = (f_e)_{e \in \binom{\sR}{2} \setminus \{q\}}$ depends only on the ring labeling and is set to zero outside $\event$.
	The coefficients satisfy $f_e = 0$ whenever $\vb_e = \vzero$ and $\|\vf\|_2^2 \le 2 \ell b$.
	Furthermore,
	\begin{align} \label{eqn:endpoint-cross}
		\E\Biggl[\, \sum_{0 \le i < j < \ell} L(i,j)^8 \sum_{\substack{u \in \sB_i, v \in \sB_j, \\ uv \ne q}} f_{uv}^2 \,\bigg|\, \event\Biggr] \le 25 \ell.
	\end{align}
\end{lemma}

In words, this lemma guarantees that, with high probability over the random ring labeling, the query column $\vb_q$ can be expressed as a linear combination of the other columns, and in expectation, the representation coefficients have bounded sum of squares weighted by the eighth power of the cyclic block distance.

The two coefficient bounds account for the different base weights in \eqref{eqn:physical-scales}.
The bound on $\|\vf\|_2^2$ suffices for edges within blocks, whose base weight is $w_{\mathrm{int}} = n^2$.
For edges between blocks, $w_{uv} = w_{\mathrm{ext}} L(i,j)^{-4}$, so their contribution $f_{uv}^2 / w_{uv}^2$ is $w_{\mathrm{ext}}^{-2} L(i,j)^8 f_{uv}^2$.
We therefore need the distance-weighted bound in \eqref{eqn:endpoint-cross}.

\subsection{A Telescoping Decomposition of the Query Column}

Let $b = t+s+1$, with $t$ and $s$ to be chosen later.
Before assigning vertex labels, designate $t$ reserve slots, $s$ auxiliary slots, and one spare slot in each block.
Write $\sT_i,\sS_i \subseteq \sB_i$ for the labels assigned by $\pi$ to the reserve and auxiliary slots of block $i$, so $|\sT_i| = t$ and $|\sS_i| = s$.
The query endpoints $x$ and $y$ occupy the spare slots of blocks $0$ and $\ell / 2$, respectively.
The slot groups are fixed in advance; only their labels are random.
These sets are used only in the analysis and do not change the hard distributions.

For each reserve $\sT_i$, let $\mB_{\sT_i}$ be the matrix of columns on edges within $\sT_i$, and let $\mB_{v, \sT_i}$ be the matrix of columns on edges joining $v \notin \sT_i$ to $\sT_i$.
Using $y$ as a common root, we call $v \notin \sT_i \cup \{y\}$ \textit{reserve-good} for $\sT_i$ if $\lev_{\vb_{vy}}\bigl([\mB_{v, \sT_i} \  \mB_{y, \sT_i} \  \mB_{\sT_i}]\bigr) \le 1$.

\begin{lemma} \label[lemma]{lem:reserve-good-prob}
	Fix a block index $i \in \{0,\dots,\ell-1\}$.
	Under the uniform ring labeling, with $\sT_i$ given by the designated reserve slots of block $i$ and $x$ and $y$ given by the two spare slots in blocks $0$ and $\ell/2$, we have
	\begin{align} \label{eqn:reserve-good-prob}
		\Pr\{x\textnormal{ is not reserve-good for }\sT_i \textnormal{ with respect to }y\} \le \frac{4N}{|\sR| (t + 1)}.
	\end{align}
\end{lemma}

\begin{proof}
	For the fixed incidence sketch matrix $\mB$, we assign each row to a vertex incident to every edge in its support (zero rows can be assigned arbitrarily).
	Note that we fix the row assignments before the random ring labeling.
	We let $\sI_v \subseteq [N]$ be the set of row indices assigned to vertex $v$, and let $k_v \coloneqq |\mathcal{I}_v|$.
	It holds that $\supp(\vb_{uv}) \subseteq \mathcal{I}_u \cup \mathcal{I}_v$ and $\sum_{v \in \sR} k_v = N$.

	Condition on the vertex set $\sT_i \cup \{x,y\}$.
	The $\vb$ columns of its induced clique have rank at most $\sum_{v \in \sT_i \cup \{x,y\}} k_v$, because they are supported on the coordinates associated with these vertices.
	As the query edge $xy$ is uniform among the $\binom{t + 2}{2}$ edge slots, \cref{lem:large_leverage_bound} upper bounds the conditional probability that the leverage score exceeds one by $2 \sum_{v \in \sT_i \cup \{x,y\}} k_v / \binom{t + 2}{2}$.
	The set $\sT_i \cup \{x,y\}$ is a uniform $(t+2)$-vertex subset of $\sR$, so $\E\bigl[\sum_{v \in \sT_i \cup \{x,y\}} k_v\bigr] = (t+2) N / |\sR|$.
	Averaging the conditional bound therefore leads to the upper bound of $2(t+2)N / \bigl(|\sR| \binom{t + 2}{2}\bigr) = 4N / \bigl(|\sR| (t+1)\bigr)$, as desired.
\end{proof}

For any $v \in \sR \setminus (\sT_i \cup \{y\})$ that is reserve-good for $\sT_i$, we write its minimum-norm representation, with coefficient coordinates ordered by vertex labels, as
\begin{align} \label{eqn:reserve-decomp}
	\vb_{vy} = \mB_{v, \sT_i} \valpha^{(v)}_{v, \sT_i} + \mB_{y, \sT_i} \valpha^{(v)}_{y, \sT_i} + \mB_{\sT_i} \valpha^{(v)}_{\sT_i},
\end{align}
where the three coefficient vectors correspond to edges from $v$ to $\sT_i$, from $y$ to $\sT_i$, and within $\sT_i$, respectively.
By reserve-goodness,
\begin{align*}
	\bigl\|\valpha^{(v)}_{v, \sT_i}\bigr\|_2^2 + \bigl\|\valpha^{(v)}_{y, \sT_i}\bigr\|_2^2 + \bigl\|\valpha^{(v)}_{\sT_i}\bigr\|_2^2 \le 1.
\end{align*}

The term $\mB_{y, \sT_i}\valpha^{(v)}_{y, \sT_i}$ uses the star from $y$ to $\sT_i$; we call it the \emph{root-star term}.
For $v = x$ and $i = 0$, the other two terms in \eqref{eqn:reserve-decomp} are represented within block $0$, as illustrated in \cref{fig:overview-reserve}.
We will replace the remaining root-star term by one using $\sT_{\ell / 2}$, which is represented within the block containing $y$.

We will choose a path $0 = i_0 < i_1 < \cdots < i_m = \ell/2$ such that $x$ is reserve-good for every selected reserve $\sT_{i_0}, \sT_{i_1}, \dots, \sT_{i_m}$.
Adding and subtracting the root-star terms along this path gives the following telescoping decomposition of the query column $\vb_{xy}$:
\begin{equation} \label{eqn:telescoping}
	\begin{aligned}
		\vb_{xy} = {} & \bigl(\vb_{xy} - \mB_{y, \sT_0} \valpha^{(x)}_{y, \sT_0}\bigr) \\
		& + \sum_{j=1}^{m} \bigl(\mB_{y, \sT_{i_{j-1}}} \valpha^{(x)}_{y, \sT_{i_{j-1}}} - \mB_{y, \sT_{i_j}} \valpha^{(x)}_{y, \sT_{i_j}}\bigr) \\
		& + \mB_{y, \sT_{\ell/2}} \valpha^{(x)}_{y, \sT_{\ell/2}}.
	\end{aligned}
\end{equation}
It remains to represent the differences in this sum and choose the path so that their total distance-weighted coefficient cost is small.

\subsection{Representing the Telescoping Terms}

For two reserves $\sT_i,\sT_j$, the differences of their root-star terms $\mB_{y, \sT_i} \valpha^{(x)}_{y, \sT_i} - \mB_{y, \sT_j} \valpha^{(x)}_{y, \sT_j}$ lie in a space of dimension at most $2t$.
We use the symmetry between $x$ and the auxiliary vertices in $\sS_i$ to represent the difference for $x$ by those for the auxiliary vertices.
Subtracting each auxiliary vertex's two reserve representations cancels its column to $y$, leaving only edges within the reserves and from the auxiliary vertex to the reserves; see \cref{fig:overview-transition}.

\begin{lemma} \label[lemma]{lem:telescope-term}
	Fix distinct block indices $i,j \in \{0,\dots,\ell-1\}$, and consider the vertex sets $\sT_i,\sT_j,\sS_i$.
	Under the uniform ring labeling, with probability at least $1 - \frac{8N}{|\sR|(t+1)} - \frac{4t}{s+1}$, the following holds:
	\begin{enumerate}
		\item $x$ is reserve-good for both $\sT_i$ and $\sT_j$, and
		\item there are coefficients $\vf = (f_e)_{e \in \binom{\sR}{2} \setminus \{q\}}$ such that
		\begin{align*}
			\mB_{y, \sT_i} \valpha^{(x)}_{y, \sT_i} - \mB_{y, \sT_j} \valpha^{(x)}_{y, \sT_j} = \sum_{e \in \binom{\sR}{2} \setminus \{q\}} f_e \vb_e,
		\end{align*}
		and furthermore, $\vf$ satisfies
		\begin{align*}
			\supp(\vf) \subseteq \binom{\sT_i}{2} \cup \binom{\sT_j}{2} \cup \{uv:u \in \sS_i, v \in \sT_i \cup \sT_j\}
		\end{align*}
		and
		\begin{align}
			\sum_{u \in \sS_i} \sum_{v \in \sT_j} f_{uv}^2 & \le 1, \label{eqn:telescope-cost_across} \\
			\sum_{e \in \binom{\sT_i}{2} \cup \binom{\sT_j}{2}} f_e^2 + \sum_{u \in \sS_i} \sum_{v \in \sT_i} f_{uv}^2 & \le 2s+1. \label{eqn:telescope-cost_internal}
		\end{align}
	\end{enumerate}
\end{lemma}

Note that for the nonzero entries in the stated coefficient vector $\vf$, only those corresponding to edges between $\sS_i$ and $\sT_j$ are across different blocks, and the sum of their squares is bounded by $1$ in \eqref{eqn:telescope-cost_across}.

\begin{proof}[Proof of \cref{lem:telescope-term}]
	For each $v \in \sS_i \cup \{x\}$ that is reserve-good for both reserves, define $\vbeta_v \coloneqq \mB_{y, \sT_i}\valpha^{(v)}_{y, \sT_i} - \mB_{y, \sT_j}\valpha^{(v)}_{y, \sT_j}$.
	For the other vertices in $\sS_i\cup\{x\}$, set $\vbeta_v \coloneqq \vzero_{N}$.
	This defines a vector for every vertex in $\sS_i\cup\{x\}$, so we can use the symmetry between the query slot $x$ and all auxiliary slots without conditioning on reserve-goodness.
	Let $\event_1$ be the event that $x$ is reserve-good for both reserves and $\lev_{\vbeta_x} \bigl((\vbeta_v)_{v \in \sS_i}\bigr) \le 1$.
	This event depends only on $x$, $y$, $\sT_i$, $\sT_j$, and $\sS_i$.

	We first lower bound $\Pr\{\event_1\}$.
	Condition on $y$, $\sT_i,\sT_j$, and the vertex set $\sS_i \cup \{x\}$.
	The vectors $(\vbeta_v)_{v \in \sS_i \cup \{x\}}$ are then fixed, and $x$ is uniform in this vertex set.
	Each $\vbeta_v$ lies in $\range\bigl([\mB_{y, \sT_i} \  \mB_{y, \sT_j}]\bigr)$, so the matrix $\bigl[\vbeta_v: v \in \sS_i \cup \{x\}\bigr]$ has rank at most $2t$.
	By \cref{lem:large_leverage_bound}, the conditional probability that $\lev_{\vbeta_x} \bigl((\vbeta_v)_{v \in \sS_i}\bigr)$ exceeds one is at most $4t/(s+1)$.
	Averaging over the conditioning and applying \cref{lem:reserve-good-prob} to both reserves, a union bound gives $\Pr\{\event_1\} \ge 1 - \frac{8N}{|\sR|(t+1)} - \frac{4t}{s+1}$. 

	On $\event_1$, choose minimum-norm coefficients $(c_v)_{v \in \sS_i}$ with $\vbeta_x = \sum_{v \in \sS_i} c_v \vbeta_v$.
	Then $\sum_{v \in \sS_i} c_v^2 \le 1$, and $c_v = 0$ whenever $\vbeta_v = \vzero$.
	For each $v$ with $c_v \ne 0$, both reserve representations exist; subtracting them cancels $\vb_{vy}$ and gives
	\begin{align*}
		\vbeta_v = \mB_{\sT_j} \valpha^{(v)}_{\sT_j} - \mB_{\sT_i} \valpha^{(v)}_{\sT_i} + \mB_{v, \sT_j} \valpha^{(v)}_{v, \sT_j} - \mB_{v, \sT_i} \valpha^{(v)}_{v, \sT_i}.
	\end{align*}
	Substituting this into $\vbeta_x = \sum_{v \in \sS_i} c_v \vbeta_v$ gives
	\begin{align*}
		\vbeta_x = {} & \sum_{v \in \sS_i, c_v \ne 0} c_v \bigl(\mB_{\sT_j} \valpha^{(v)}_{\sT_j} - \mB_{\sT_i} \valpha^{(v)}_{\sT_i} + \mB_{v, \sT_j} \valpha^{(v)}_{v, \sT_j} - \mB_{v, \sT_i} \valpha^{(v)}_{v, \sT_i}\bigr) \\
		= {} & \mB_{\sT_j} \sum_{v \in \sS_i, c_v \ne 0} c_v \valpha^{(v)}_{\sT_j} - \mB_{\sT_i} \sum_{v \in \sS_i, c_v \ne 0} c_v \valpha^{(v)}_{\sT_i} \\
		& + \sum_{v \in \sS_i, c_v \ne 0} \mB_{v, \sT_j} \bigl(c_v \valpha^{(v)}_{v, \sT_j}\bigr) - \sum_{v \in \sS_i, c_v \ne 0} \mB_{v, \sT_i} \bigl(c_v \valpha^{(v)}_{v, \sT_i}\bigr).
	\end{align*}
	Reading off the coefficient of each column, define $\vf = (f_e)_{e \in \binom{\sR}{2} \setminus \{q\}}$ by
	\begin{align*}
		f_e \coloneqq
		\begin{cases}
			- \displaystyle \sum_{v \in \sS_i, c_v \ne 0} c_v \bigl(\valpha^{(v)}_{\sT_i}\bigr)_e, & e \in \binom{\sT_i}{2}, \\
			\displaystyle \sum_{v \in \sS_i, c_v \ne 0} c_v \bigl(\valpha^{(v)}_{\sT_j}\bigr)_e, & e \in \binom{\sT_j}{2}, \\
			-c_v \bigl(\valpha^{(v)}_{v, \sT_i}\bigr)_{uv}, & e = uv, u \in \sT_i, v \in \sS_i, c_v \ne 0, \\
			c_v \bigl(\valpha^{(v)}_{v, \sT_j}\bigr)_{uv}, & e = uv, u \in \sT_j, v \in \sS_i, c_v \ne 0, \\
			0, & \text{otherwise}.
		\end{cases}
	\end{align*}
	Hence we have $\vbeta_x = \sum_{e \in \binom{\sR}{2} \setminus \{q\}} f_e \vb_e$ and $\supp(\vf) \subseteq \binom{\sT_i}{2} \cup \binom{\sT_j}{2} \cup \{uv:u \in \sS_i, v \in \sT_i \cup \sT_j\}$.
	We next bound the sum of squares of the coefficients.
	By the property of reserve-goodness, we have
	\begin{align*}
		\sum_{v \in \sS_i} \sum_{u \in \sT_j} f_{vu}^2 & = \sum_{v \in \sS_i, c_v \ne 0} c_v^2 \, \bigl\|\valpha^{(v)}_{v, \sT_j}\bigr\|_2^2 \le \sum_{v \in \sS_i} c_v^2 \le 1.
	\end{align*}
	For the remaining coefficients, we have
	\begin{align*}
		& \sum_{e \in \binom{\sT_i}{2} \cup \binom{\sT_j}{2}} f_e^2 + \sum_{v \in \sS_i} \sum_{u \in \sT_i} f_{vu}^2 \\
		= {} & \biggl\|\sum_{v \in \sS_i, c_v \ne 0} c_v \valpha^{(v)}_{\sT_i}\biggr\|_2^2 + \biggl\|\sum_{v \in \sS_i, c_v \ne 0} c_v \valpha^{(v)}_{\sT_j}\biggr\|_2^2 + \sum_{v \in \sS_i, c_v \ne 0} c_v^2 \, \bigl\|\valpha^{(v)}_{v, \sT_i}\bigr\|_2^2 \\
		\le {} & \biggl(\sum_{v \in \sS_i} c_v^2\biggr) \sum_{v \in \sS_i, c_v \ne 0} \Bigl(\bigl\|\valpha^{(v)}_{\sT_i}\bigr\|_2^2 + \bigl\|\valpha^{(v)}_{\sT_j}\bigr\|_2^2\Bigr) + 1 \le 2s+1,
	\end{align*}
	where we used the Cauchy--Schwarz inequality and $|\sS_i| = s$.
	This completes the proof.
\end{proof}

\subsection{Low-Cost Routing on the Block Cycle}

For any given integer $t \ge 1$, set $s \coloneqq 10^8 t$ and $b \coloneqq t + s + 1$.
We now choose the block indices in \eqref{eqn:telescoping}.
For distinct block indices $i,j$, call $i\to j$ a \emph{transition} if the event $\event_1$ in the proof of \cref{lem:telescope-term} holds for this ordered pair.
On this event, $x$ is reserve-good for both reserves $\sT_i,\sT_j$, and the lemma represents the difference of their root-star terms using $\sS_i$, with the stated coefficient bounds.
The transition depends only on $x,y,\sT_i,\sT_j,\sS_i$ and is directed as it only uses the auxiliary set of the source block.

\begin{lemma}[Routing with bounded expected cost] \label[lemma]{lem:short-transition-routing}
	Let $\ell \ge 10$ be even, with $t,s,b$ as above.
	Fix any $\sR$ of size $\ell b$ and an incidence sketch matrix $\mB = \bigl[\vb_e: e \in \binom{\sR}{2}\bigr]$ of dimension $N \le 10^{-9} \ell b t$.
	Under the uniform ring labeling, there is an event $\event$ with $\Pr\{\event\} \ge 0.999$ on which we can choose an increasing path
	\begin{align*}
		0 = i_0 < i_1 < \cdots < i_m = \ell / 2
	\end{align*}
	of transitions.
	The path is determined by the labeling and satisfies
	\begin{align*}
		\E\Bigl[\sum_{j=1}^{m} L(i_j,i_{j-1})^8 \, \bigm| \, \event\Bigr] \le 25 \ell.
	\end{align*}
\end{lemma}

The cost in this lemma comes from the distance-weighted sum in \eqref{eqn:endpoint-cross}.
The cross-block weights in \eqref{eqn:physical-scales} are proportional to $L(i,j)^{-4}$, so their inverse squares contribute a factor of $L(i,j)^8$.
For a transition $i\to j$ with $0\le i < j \le \ell/2$, the only cross-block coefficients are on edges between $\sS_i$ and $\sT_j$, all at cyclic distance $L(i,j) = j-i$.
By \eqref{eqn:telescope-cost_across}, their squared sum is at most one, so their contribution to the distance-weighted sum satisfies
\begin{align*}
	L(i,j)^8 \sum_{u \in \sS_i} \sum_{v \in \sT_j} f_{uv}^2 = (j-i)^8 \sum_{u \in \sS_i} \sum_{v \in \sT_j} f_{uv}^2 \le (j-i)^8.
\end{align*}

\begin{proof}
	Call a block $i$ \emph{usable} if, among all $\ell-1$ other blocks $j$, at most $(\ell-1) / 100$ lack the transition $i \to j$ and at most $(\ell-1) / 100$ lack the transition $j \to i$.
	Thus a usable block has few missing arrows in either direction.
	Let $\event$ be the event that fewer than $\ell / 100$ blocks are unusable and both blocks $0$ and $\ell / 2$ are usable.

	For any fixed block $i$, by \cref{lem:telescope-term} and linearity of expectation, we have the following bound on the expected number of missing incoming and outgoing arrows.
	\begin{align*}
		\E\Bigl[\bigl|\{j \ne i: i \nrightarrow j\}\bigr| + \bigl|\{j \ne i: j \nrightarrow i\}\bigr|\Bigr] \le 2 (\ell-1) \biggl(\frac{8N}{|\sR|(t+1)} + \frac{4t}{s+1}\biggr).
	\end{align*}
	If block $i$ is unusable, this total exceeds $(\ell-1)/100$.
	Markov's inequality therefore gives
	\begin{align*}
		\Pr\{i \text{ is unusable}\} \le 200\left(\frac{8N}{|\sR|(t+1)} + \frac{4t}{s+1}\right) \le 200 \cdot 4.8 \cdot 10^{-8},
	\end{align*}
	where we used $N \le 10^{-9} |\sR| t$ and $s = 10^8 t$.
	Summing over blocks bounds the expected number of unusable blocks by $200 \cdot 4.8 \cdot 10^{-8} \ell$.
	Markov's inequality bounds the probability of having at least $\ell / 100$ unusable blocks by $100 \cdot 200 \cdot 4.8 \cdot 10^{-8}$.
	Adding the failure probabilities of the two endpoint blocks gives
	\begin{align*}
		1 - \Pr\{\event\} \le (100+2) \cdot 200 \cdot 4.8 \cdot 10^{-8} < 0.001.
	\end{align*}

	On $\event$, an intermediate block $i$ gives a path $0 \to i \to \ell / 2$ if it is usable and both arrows are present.
	Fewer than $\ell / 100$ blocks are unusable.
	Since blocks $0$ and $\ell / 2$ are usable, at most $(\ell-1) / 100$ intermediate blocks $i$ lack the arrow $0 \to i$, and at most $(\ell-1)/100$ lack the arrow $i \to \ell / 2$.
	Thus fewer than $3 \ell / 100 < \ell / 2 - 1$ intermediate positions are excluded, so such a path exists.

	Among all increasing paths from $0$ to $\ell / 2$ through usable blocks and transitions, choose one that visits the maximum number of blocks, breaking ties arbitrarily.
	This choice prevents inserting any further usable block between consecutive path indices with both required arrows present.

	Only a small fraction of blocks fail insertion, so the random order makes large gaps between consecutive path indices unlikely.
	We use this to bound the expected sum of their costs $(j-i)^8$.

	\paragraph{Probability of a selected gap.}
	First condition on the query endpoints $x,y$ and the unordered collection of pairs $\bigl\{(\sT_i,\sS_i): 0 \le i < \ell \bigr\}$.
	This fixes how the vertices are grouped into reserve--auxiliary pairs, but not how those pairs are ordered on the ring.
	The arrows depend only on $x,y$ and these vertex sets, so all arrows between pairs and the usability of each pair are fixed.
	The order of the pairs remains uniform as permuting whole pairs while keeping all spare-slot labels unchanged gives a bijection between labelings with any two prescribed orders.
	Next fix the pairs at positions $0,\ell/2$.
	These choices determine whether $\event$ holds as we know the number of unusable pairs and whether the two endpoint pairs are usable.
	Consider any such choice for which $\event$ holds.
	Under this further conditioning, the remaining pairs are still uniformly ordered.

	For fixed positions $0 \le i < j \le \ell / 2$, let $X_{ij}$ indicate that the selected path uses the step $i \to j$.
	Further fix the reserve--auxiliary pairs at $i,j$, without conditioning on whether $X_{ij}=1$.
	If either pair is unusable or the arrow $i\to j$ is missing, then $X_{ij} = 0$.
	Otherwise, suppose the selected path uses this step.
	Any usable block $k$ with $i < k < j$ and $i \to k \to j$ could be inserted while keeping all other path indices unchanged.
	The resulting path would visit one more block, contradicting the choice of the path.
	Thus $X_{ij} = 1$ requires every pair strictly between $i$ and $j$ to fail at least one insertion condition.
	There are fewer than $\ell / 100$ unusable pairs, and the usable pairs at $i,j$ each exclude at most $(\ell-1) / 100$ others through a missing required arrow.
	Thus fewer than $3\ell / 100$ other pairs fail an insertion condition.
	The set of these pairs is fixed under the current conditioning.

	At most four pairs have been fixed, at positions $0,\ell / 2,i,j$, none strictly between $i$ and $j$.
	The $j-i-1$ interior pairs are a uniform sample without replacement from at least $\ell-4$ remaining pairs.
	Hence the conditional probability that all of them fail an insertion condition is at most
	\begin{align*}
		\Bigl(\frac{3\ell}{100(\ell-4)}\Bigr)^{j-i-1} \le 20^{-(j-i-1)},
	\end{align*}
	where $\ell \ge 10$.
	Conditional on every previous draw failing an insertion condition, the fraction of remaining pairs that fail a condition can only decrease.
	Multiplying these conditional bounds gives the displayed inequality; if there are too few such pairs to fill the interval, the probability is zero.
	This bound holds for every choice of the fixed data consistent with $\event$, so averaging over these data gives $\E[X_{ij} \mid \event] \le 20^{-(j-i-1)}$.

	\paragraph{Expected cost.}
	On $\event$, every step of the selected path contributes to exactly one indicator $X_{ij}$.
	Thus, 
	\begin{align*}
		\sum_{a=1}^{m} L(i_a,i_{a-1})^8 = \sum_{0 \le i < j \le \ell/2} (j-i)^8 X_{ij}.
	\end{align*}
	Taking conditional expectations, applying the bound on $\E[X_{ij}\mid\event]$, and grouping the pairs by $d=j-i$, we obtain
	\begin{align*}
		\E\Biggl[\sum_{a=1}^{m} L(i_a,i_{a-1})^8 \, \bigg| \, \event\Biggr] & = \sum_{0 \le i < j \le \ell / 2} (j-i)^8 \, \E[X_{ij} \mid \event] \\
		& \le \sum_{0 \le i < j \le \ell / 2} (j-i)^8 \, 20^{-(j-i-1)} \\
		& = \sum_{d=1}^{\ell/2} \sum_{i=0}^{\ell/2 - d} d^8 \, 20^{-(d-1)} \\
		& = \sum_{d=1}^{\ell/2} (\ell/2 + 1 - d) \, d^8 \, 20^{-(d-1)} \\
		& \le \frac{\ell}{2} \sum_{d=1}^{\ell/2} d^8 \, 20^{-(d-1)} \le \frac{\ell}{2} \sum_{d \ge 1} d^8 \, 20^{-(d-1)}.
	\end{align*}
	Here $\ell / 2 + 1 - d$ counts the pairs of positions at distance $d$.
	From $d = 3$ onward, the ratio of consecutive summands is at most $(4/3)^8 / 20 < 1/2$, so
	\begin{align*}
		\sum_{d \ge 1} d^8 \, 20^{-(d-1)} \le 1 + \frac{256}{20} + \frac{2 \cdot 6561}{400} < 50.
	\end{align*}
	It follows that $\E\bigl[\sum_{j=1}^{m} L(i_j,i_{j-1})^8 \mid \event\bigr] \le 25 \ell$, as claimed.
\end{proof}

\begin{proof}[Proof of \cref{lem:endpoint-representation}]
	Let $\event$ be the success event from \cref{lem:short-transition-routing}, so $\Pr\{\event\} \ge 0.999$.
	On $\event$, take the path $0 = i_0 < \cdots < i_m = \ell/2$ supplied by that lemma.
	By the reserve decomposition \eqref{eqn:reserve-decomp}, the first and last terms in \eqref{eqn:telescoping} have representations within blocks $0$ and $\ell / 2$, respectively, each with squared coefficient norm at most one.

	Apply \cref{lem:telescope-term} to each transition in \eqref{eqn:telescoping}.
	Summing the resulting coefficient vectors with those for the two endpoint terms gives $\vf$ satisfying
	\begin{align*}
		\vb_q = \sum_{e \in \binom{\sR}{2} \setminus \{q\}} f_e \vb_e.
	\end{align*}
	Set $f_e = 0$ whenever $\vb_e = \vzero$.

	\paragraph{Squared coefficient norm.}
	An edge within a reserve occurs in at most two representations: those for the entering and leaving transitions, or, at block $0$, the source term and the first transition.
	Every other edge occurs at most once, since the reserve and auxiliary sets are disjoint and $x,y$ occupy spare slots.
	The path is increasing, so it has $m \le \ell/2$ transitions.
	Applying $(a+b)^2 \le 2a^2 + 2b^2$ to overlapping coefficients and using the bounds in \cref{lem:telescope-term} gives
	\begin{align*}
		\|\vf\|_2^2 \le 2 \bigl(2 + m (2s+2)\bigr) \le 4 + 2 \ell(s+1) \le 2 \ell b.
	\end{align*}
	The last inequality uses $b = t+s+1$ and $2 \ell t \ge 4$.

	\paragraph{Distance-weighted coefficient sum.}
	The cross-block edges used by distinct transitions are disjoint: each joins the source auxiliary set to the destination reserve, and the source auxiliary sets are distinct along the increasing path.
	For transition $j$, their cyclic block distance is $L(i_j,i_{j-1}) = i_j - i_{j-1}$, since the path lies in the first half of the ring.
	Combining this with \eqref{eqn:telescope-cost_across} gives
	\begin{align*}
		\sum_{0 \le i < j < \ell} L(i,j)^8 \sum_{\substack{u \in \sB_i, v \in \sB_j, \\ uv \ne q}} f_{uv}^2 \le \sum_{j=1}^{m} L(i_j,i_{j-1})^8.
	\end{align*}
	This comparison holds for every labeling in $\event$.
	Taking the conditional expectation and applying \cref{lem:short-transition-routing} proves \eqref{eqn:endpoint-cross}.
	The coefficients are determined by the labeling because the coordinate orders and tie-breaking rule are fixed; set $\vf = \vzero$ off $\event$.
\end{proof}

\subsection{Proving the Lower Bound for Incidence Sketches}

\begin{proof}[Proof of \cref{thm:incidence_sketch}]
	Choose an arbitrary integer $t \ge 1$ and set $b \coloneqq \bigl(10^8 + 1\bigr)t+1$, $\ell \coloneqq 10^7 b^4$, and $n \coloneqq \ell b = \Theta\bigl(t^5\bigr)$.
	Suppose the sketch has dimension $N \le 10^{-9} nt$.
	Note that $nt = \Theta\bigl(n^{6/5}\bigr)$.
	By \cref{lem:reduction-to-leverage}, it thus suffices to prove that, for every such fixed incidence sketch matrix $\mB$,
	\begin{align*}
		\E_{\pi}\Bigl[ \min\Bigl\{ 1, \lev_{\vb_q}\Bigl(\bigl(\mB \diag(\vw)\bigr)_{-q}\Bigr) \Bigr\} \Bigr] < \frac{1}{100}.
	\end{align*}

	Let $\event$ and $\vf = (f_e)_{e \in \binom{\sR}{2} \setminus \{q\}}$ be the success event and coefficient vector supplied by \cref{lem:endpoint-representation}.
	Then $\Pr\{\event\} \ge 0.999$, and $\vf = \vzero$ whenever $\event$ does not occur.
	On $\event$, these coefficients represent $\vb_q$ using the other columns and satisfy $\|\vf\|_2^2 \le 2 \ell b$.
	By the definition of leverage score, on this event we have
	\begin{align*}
		\lev_{\vb_q}\Bigl(\bigl(\mB \diag(\vw)\bigr)_{-q}\Bigr) \le \sum_{e \in \binom{\sR}{2} \setminus \{q\}} \frac{f_e^2}{w_e^2} \le \frac{2}{\ell^3 b^3} + \sum_{0 \le i < j < \ell} \, \sum_{\substack{u \in \sB_i, v \in \sB_j, \\ uv \ne q}} \frac{f_{uv}^2}{w_{uv}^2},
	\end{align*}
	where the second inequality uses $w_{\mathrm{int}} = \ell^2 b^2$ and $\|\vf\|_2^2 \le 2 \ell b$ to bound the within-block contribution.
	Since $\vf = \vzero$ outside $\event$, the cross-block base weights and \cref{lem:endpoint-representation} give
	\begin{align}
		& \E_{\pi}\Biggl[\sum_{0 \le i < j < \ell} \, \sum_{\substack{u \in \sB_i, v \in \sB_j, \\ uv \ne q}} \frac{f_{uv}^2}{w_{uv}^2}\Biggr] \notag \\
		= {} & \frac{\Pr\{\event\}}{w_{\mathrm{ext}}^2} \E_{\pi}\Biggl[\, \sum_{0 \le i < j < \ell} L(i,j)^8 \sum_{\substack{u \in \sB_i, v \in \sB_j, \\ uv \ne q}} f_{uv}^2 \,\bigg|\, \event\Biggr] \le \frac{25 \ell}{w_{\mathrm{ext}}^2} = \frac{6400 \, b^4}{\ell}. \label{eqn:incidence-cross-energy}
	\end{align}

	If $\event$ fails, we use the trivial bound $1$ on the truncated leverage score.
	Thus, combining this failure probability with the above bounds, we have
	\begin{align*}
		\E_{\pi}\Bigl[ \min\Bigl\{ 1, \lev_{\vb_q}\Bigl(\bigl(\mB \diag(\vw)\bigr)_{-q}\Bigr) \Bigr\} \Bigr] \le \bigl(1 - \Pr\{\event\}\bigr) + \frac{2}{\ell^3 b^3} + \frac{6400 \, b^4}{\ell} < \frac{1}{100},
	\end{align*}
	where the last inequality follows from $\ell = 10^7 b^4$ and $b \ge 10^8$.

	Finally, by padding the hard-instance graphs with isolated vertices, we can extend the lower bound to all sufficiently large $n$.
	This completes the proof of \cref{thm:incidence_sketch}.
\end{proof}

\section{Lower Bound for Arbitrary Linear Sketches} \label{sec:arbitrary-sketch}

This section proves the lower bound for arbitrary linear sketches, as stated in \cref{thm:main}.

\subsection{Adding an Anchor Clique to the Hard Distributions}

We add an \emph{anchor clique} to the ring construction of \cref{sec:hard-distributions}, with no edges between the clique and the ring.
The anchor columns let us reduce the query-column representation problem to an incidence sketch on the ring.
We take the clique comparable in size to the ring to keep the expected dimension of this incidence sketch within a constant factor of the original sketch dimension.
The large anchor weights keep the additional anchor contribution to leverage small.

Given an integer $b \ge 2$ and an even integer $\ell \ge 2$, set $n \coloneqq 2 \ell b$.
Here $n$ counts both the ring and anchor vertices.
Take a template consisting of $\ell b$ anchor slots and $\ell$ ring blocks of $b$ slots each, with the same two distinguished query slots as in \Cref{sec:hard-distributions}.
Number these $n$ slots in a fixed order, and place vertex $\pi(t)$ in slot $t$, where $\pi$ is a uniformly random permutation of $[n]$.
Write $\sC$ for the anchor vertex set, $\sR \coloneqq [n] \setminus \sC$ for the ring vertex set, and $\sB_0,\dots,\sB_{\ell-1}$ for the ring blocks.
Thus $|\sC| = |\sR| = \ell b$.
The query edge is $q \coloneqq xy$, where $x,y$ occupy the distinguished slots in blocks $0,\ell/2$.

We still use $w_{\mathrm{ext}} \coloneqq \ell / \bigl(16 b^2\bigr)$, but set $w_{\mathrm{int}} \coloneqq n^2$ using the enlarged vertex count.
The edge set and base weights are now
\begin{align}
	E & \coloneqq \binom{\sC}{2} \cup \binom{\sR}{2}, \notag \\
	w_e & \coloneqq
	\begin{cases}
		w_{\mathrm{int}}, & e \in \binom{\sC}{2} \text{ or } e \in \binom{\sB_i}{2} \text{ for some } i, \\
		w_{\mathrm{ext}} \, L(i,j)^{-4}, & e = uv, u \in \sB_i, v \in \sB_j, i \ne j, \\
		0, & \text{otherwise.}
	\end{cases} \label{eqn:anchored-scales}
\end{align}
Conditional on $\pi$, independently sample $Z_e$ with density \eqref{eqn:beta-density} for each $e \in E$.
Independently choose $\Yes$ or $\No$, each with probability $1/2$.
In the $\No$ case, set $\tw_e \coloneqq w_e Z_e$ for $e \in E$; in the $\Yes$ case, set $\tw_e \coloneqq w_e Z_e + \mathbf{1}\{e=q\}$.
All coordinates outside $E$ are zero in both cases.
Throughout this section, $\Yes$ and $\No$ refer to these distributions, and $\tvw$ denotes the resulting input weight vector.

Every graph in either distribution has positive-weight ratio at most $4n^5$, by the same calculation as in the proof of \cref{lem:weight-range}.
Indeed, \eqref{eqn:anchored-scales} gives $1/n^3 \le 1/(\ell^3 b^2) \le w_e \le n^2$ for every $e \in E$.
Since $Z_e \in [1/2,3/2]$ and the query-edge weight may increase by one, every positive input weight lies in $[1/(2n^3),2n^2]$.

The next lemma shows that the leverage-score reduction in \cref{lem:reduction-to-leverage} remains valid after adding the anchor clique.

\begin{lemma} \label[lemma]{lem:anchored-reduction-to-leverage}
	If a linear sketch satisfies
	\begin{align}
		\E_{\pi}\Bigl[ \min\Bigl\{ 1, \lev_{\va_q}\Bigl(\bigl(\mA \diag(\vw)\bigr)_{-q}\Bigr) \Bigr\} \Bigr] < \frac{1}{100}
	\end{align}
	for every realization $\mA$ of its sketch matrix, with the base weights in \eqref{eqn:anchored-scales}, then no recovery algorithm for this sketch can compute a $(1 \pm 0.01)$-spectral approximation with probability at least $0.99$ on every input in the support of either anchored distribution.
\end{lemma}

\begin{proof}
	We check that the two ingredients in the proof of \cref{lem:reduction-to-leverage} remain valid.
	Extend the potential $\vphi$ from the proof of \cref{lem:spectral-separation} by zero on $\sC$.
	Anchor and within-block edges contribute zero to its quadratic form because their endpoints have equal potential.
	There are no edges between $\sC$ and $\sR$, and the cross-block weights and the unit increase on $q$ are unchanged.
	The spectral separation from \cref{lem:spectral-separation} therefore still holds, and the distinguishing algorithm constructed in its proof remains correct whenever recovery succeeds.

	Fix a realization $\mA$ of the sketch matrix and condition on $\pi$.
	Coordinates outside $E$ contribute nothing to the sketch and correspond to zero columns of $\mA\diag(\vw)$, so deleting them leaves the leverage score unchanged.
	On $E$, all anchor and ring noises are independent, and $\Yes$ differs from $\No$ only by the unit increase on $q$.
	The proof of \cref{lem:TV-leverage} therefore gives the same TV bound after restricting to these coordinates.

	Averaging over $\pi$ as in the proof of \cref{lem:reduction-to-leverage}, the leverage assumption gives TV distance less than $1/\sqrt{10}$ between $(\pi,\mA\tvw)_{\Yes}$ and $(\pi,\mA\tvw)_{\No}$ for every fixed $\mA$.
	Since the two cases are equally likely, \eqref{eqn:TV-distinguishing} gives success probability less than $3/4$ for any randomized distinguishing algorithm given $\mA$, the sketch, and $\pi$.
	This bound also holds after averaging over the independently sampled sketch matrix.

	Suppose now that a recovery algorithm satisfies the stated guarantee.
	By the spectral separation above, applying the distinguishing algorithm to its output and $\pi$ is correct whenever recovery succeeds.
	Averaging over inputs therefore gives success probability at least $0.99$ under each distribution, a contradiction.
\end{proof}

\subsection{Representing Arbitrary Sketches via Incidence Sketches}

For any realization of $\sC$ and $\sR$, we will represent the arbitrary sketch matrix $\mA$ via an incidence sketch matrix $\mB = \bigl[\vb_e: e \in \binom{\sR}{2}\bigr]$ on $\sR$, meaning that each row of $\mB$ is supported on edges in $\binom{\sR}{2}$ that are incident on a single vertex in $\sR$ (which could be different for different rows).

For any realization of $\sC$ and $\sR$, we define $\mA_{\sC} \coloneqq \bigl[\va_e: e \in \binom{\sC}{2}\bigr]$ and $\mA_{v, \sC} \coloneqq [\va_{vw}: w \in \sC]$ for $v \in \sR$.
We call an edge $uv \in \binom{\sR}{2}$ \textit{anchor-good} if $\lev_{\va_{uv}} \bigl([\mA_{u, \sC} \  \mA_{v, \sC} \  \mA_{\sC}]\bigr) \le 1$.
Using the uniformity of the query edge $q$ conditional on $\sC \cup \{x,y\}$, we immediately obtain the following bound on the probability that $q$ is not anchor-good.

\begin{lemma} \label[lemma]{lem:anchor-good-prob}
	For any fixed sketch matrix $\mA$, we have
	\begin{align} \label{eqn:anchor-good-prob}
		\Pr\nolimits_{\pi}\{q \textup{ is not anchor-good}\} \le 2 N \Big/ \binom{\ell b + 2}{2}.
	\end{align}
\end{lemma}

\begin{proof}
	Conditional on the vertex set $\sC \cup \{x,y\}$, the query edge $q = xy$ is a uniform edge of its clique, so applying \cref{lem:large_leverage_bound} gives
	\begin{equation*}
		\Pr\nolimits_{\pi}\Bigl\{\lev_{\va_q} \bigl([\mA_{x, \sC} \  \mA_{y, \sC} \  \mA_{\sC}]\bigr) > 1\Bigr\} \le 2 \rank(\mA) \Big/ \binom{|\sC| + 2}{2} \le 2 N \Big/ \binom{\ell b + 2}{2}. \qedhere
	\end{equation*}
\end{proof}

To represent arbitrary sketches via incidence sketches, we first introduce some additional notation that will be used to define the dimension of the incidence sketch.
For any realization of $\sC$, we define $k_v(\sC) \coloneqq n_{+}\bigl(\mA_{v, \sC} \mA_{v, \sC}^{\top} - \mA_{\sC} \mA_{\sC}^{\top}\bigr)$ for $v \in \sR$, where $n_{+}(\mM)$ denotes the number of positive eigenvalues of a symmetric matrix $\mM$.
We then define $K(\sC) \coloneqq \sum_{v \in \sR} k_v(\sC)$.

The next lemma shows that every column of the sketch matrix $\mA$ corresponding to an anchor-good ring edge can be decomposed into a linear transformation of the corresponding column of an incidence sketch on $\sR$ with dimension $K(\sC)$ plus a linear combination of the anchor columns.

\begin{lemma} \label[lemma]{lem:incidence-decomp}
	Consider any realization of $\sC$ and $\sR$ and any sketch matrix $\mA$.
	There exist an incidence sketch matrix $\mB = \bigl[\vb_e: e \in \binom{\sR}{2}\bigr]$ on $\sR$ with dimension $K(\sC)$, a matrix $\mT \in \R^{N \times K(\sC)}$, and vectors $(\vc_e)_{e \in \binom{\sR}{2}}$ in $\R^{\binom{\ell b}{2}}$ all determined by $\mA$ and $\sC$ that satisfy the following:
	\begin{enumerate}
		\item for every anchor-good ring edge $e$, $\va_e = \mT \, \vb_e + \mA_{\sC} \vc_e$ and $\|\vc_e\|_2^2 \le 3$;
		\item for every non-anchor-good ring edge $e$, $\vb_e = \vzero_{K(\sC)}$ and $\vc_e = \vzero_{\binom{\ell b}{2}}$.
	\end{enumerate}
\end{lemma}

\begin{proof}
	For any ring vertex $v$, let $\mT_v \in \R^{N \times k_v(\sC)}$ have the orthonormal eigenvectors corresponding to the positive eigenvalues of $\mA_{v, \sC}\mA_{v, \sC}^{\top} - \mA_{\sC}\mA_{\sC}^{\top}$ as its columns, each scaled by the square root of its eigenvalue.
	By the eigendecomposition, this guarantees that $\mA_{v, \sC} \mA_{v, \sC}^{\top} - \mA_{\sC} \mA_{\sC}^{\top} \preceq \mT_v \mT_v^{\top}$, and thus we have
	\begin{align}
		\mA_{v, \sC} \mA_{v, \sC}^{\top} \preceq \mT_v \mT_v^{\top} + \mA_{\sC} \mA_{\sC}^{\top}. \label{eqn:star-dominance}
	\end{align}
	Fix an ordering $v_1,\dots,v_{\ell b}$ of vertices in $\sR$ by vertex labels and concatenate all the matrices as $\mT \coloneqq [\mT_{v_1} \  \cdots \  \mT_{v_{\ell b}}] \in \R^{N \times K(\sC)}$.
	Let $\sI_v$ be the set of the $k_v(\sC)$ coordinates occupied by $\mT_v$.

	For any non-anchor-good ring edge $uv$, we can simply set $\vb_{uv} \coloneqq \vzero$ and $\vc_{uv} \coloneqq \vzero$ as required.
	Now consider any anchor-good ring edge $uv$.
	By the definition of anchor-goodness, it holds that $\lev_{\va_{uv}}\bigl([\mA_{u, \sC} \  \mA_{v, \sC} \  \mA_{\sC}]\bigr) \le 1$, so by \cref{lem:leverage-Loewner},
	\begin{align*}
		\va_{uv} \va_{uv}^{\top} & \preceq \mA_{u, \sC} \mA_{u, \sC}^{\top} + \mA_{v, \sC} \mA_{v, \sC}^{\top} + \mA_{\sC} \mA_{\sC}^{\top} \\
		& \preceq \mT_u \mT_u^{\top} + \mT_v \mT_v^{\top} + 3 \mA_{\sC} \mA_{\sC}^{\top},
	\end{align*}
	where we plug in \eqref{eqn:star-dominance} for both $u$ and $v$.
	Now, by \cref{lem:leverage-Loewner}, $\lev_{\va_{uv}}\bigl([\mT_u \  \mT_v \  \sqrt{3} \mA_{\sC}]\bigr) \le 1$, so there exist $\vbeta_u \in \R^{k_u(\sC)}$, $\vbeta_v \in \R^{k_v(\sC)}$, and $\vc_{uv} \in \R^{\binom{\ell b}{2}}$ such that
	\begin{align*}
		\va_{uv} = \mT_u \vbeta_u + \mT_v \vbeta_v + \mA_{\sC} \vc_{uv}
	\end{align*}
	and $\|\vbeta_u\|_2^2 + \|\vbeta_v\|_2^2 + \frac{1}{3} \|\vc_{uv}\|_2^2 \le 1$.
	In light of this, define $\vb_{uv} \in \R^{K(\sC)}$ by placing $\vbeta_u$ in $\sI_u$ and $\vbeta_v$ in $\sI_v$, with all other coordinates zero.
	This guarantees that $\supp(\vb_{uv}) \subseteq \sI_u \cup \sI_v$ and thus $\mB$ is an incidence sketch matrix on $\sR$ with dimension $K(\sC)$.
	Then, we have $\va_{uv} = \mT \, \vb_{uv} + \mA_{\sC} \, \vc_{uv}$ by construction and $\|\vc_{uv}\|_2^2 \le 3$, as desired.
\end{proof}

As a consequence of this endpoint decomposition scheme, the next lemma shows that, if $\vb_q$ can be expressed as a linear combination of other $\vb$ columns with small coefficients, then $\va_q$ can also be expressed by other $\va$ columns with small coefficients.

\begin{lemma}[Lifting] \label[lemma]{lem:lifting}
	Fix $\sC$ and $\sR$, and let $\mB = [\vb_e: e\in\binom{\sR}{2}]$ be the incidence sketch matrix supplied by \cref{lem:incidence-decomp}.
	Let $q \in \binom{\sR}{2}$ be any anchor-good ring edge.
	Suppose that coefficients $\vf = (f_e)_{e \in \binom{\sR}{2} \setminus \{q\}}$ satisfy $\vb_q = \sum_{e \in \binom{\sR}{2} \setminus \{q\}} f_e \vb_e$ and $f_e = 0$ whenever $\vb_e = \vzero$.
	Then there exists a vector $\vc \in \R^{\binom{\ell b}{2}}$ that satisfies
	\begin{align} \label{eqn:lifted-representation}
		\va_q = \sum_{e \in \binom{\sR}{2} \setminus \{q\}} f_e \va_e + \mA_{\sC} \vc
	\end{align}
	and
	\begin{align*}
		\|\vc\|_2^2 \le 6 \bigl(1 + \ell^2 b^2 \|\vf\|_2^2\bigr).
	\end{align*}
\end{lemma}

\begin{proof}
	Letting $\mT$ and $(\vc_e)$ be the matrix and vectors accompanying $(\vb_e)$ in \cref{lem:incidence-decomp}, we have $\va_e = \mT \, \vb_e + \mA_{\sC} \vc_e$ for every anchor-good ring edge $e$.
	As $\vb_e = \vzero$ for any non-anchor-good ring edge $e$, by the requirement for $\vf$ in the lemma statement, $f_e \ne 0$ implies that $e$ is anchor-good.
	Thus, we have
	\begin{align*}
		\sum_{e \in \binom{\sR}{2} \setminus \{q\}} f_e \va_e & = \mT \sum_{e \in \binom{\sR}{2} \setminus \{q\}} f_e \vb_e + \mA_{\sC} \sum_{e \in \binom{\sR}{2} \setminus \{q\}} f_e \vc_e \\
		& = \mT \, \vb_q + \mA_{\sC} \sum_{e \in \binom{\sR}{2} \setminus \{q\}} f_e \vc_e.
	\end{align*}
	As $q$ is anchor-good, we have $\va_q = \mT \, \vb_q + \mA_{\sC} \vc_q$.
	Subtracting the two equations cancels the $\mT \, \vb_q$ term,	and rearranging yields
	\begin{align*}
		\va_q = \sum_{e \in \binom{\sR}{2} \setminus \{q\}} f_e \va_e + \mA_{\sC} \biggl(\vc_q - \sum_{e \in \binom{\sR}{2} \setminus \{q\}} f_e \vc_e\biggr).
	\end{align*}
	Now let $\vc \coloneqq \vc_q - \sum_{e \in \binom{\sR}{2} \setminus \{q\}} f_e \vc_e$ so \cref{eqn:lifted-representation} holds.
	Since $\|\vc_e\|_2^2 \le 3$ for every ring edge $e$, applying the triangle inequality and the Cauchy--Schwarz inequality gives
	\begin{align*}
		\|\vc\|_2^2 \le 3 \biggl(1 + \sum_{e \in \binom{\sR}{2} \setminus \{q\}}|f_e|\biggr)^2 \le 6 \bigl(1 + \ell^2 b^2 \|\vf\|_2^2\bigr),
	\end{align*}
	which completes the proof.
\end{proof}

We next show that $\E\bigl[K(\sC)\bigr]$, the expected dimension of the incidence sketch $\mB$ constructed in \cref{lem:incidence-decomp}, is at most $4 N$.

\begin{lemma} \label[lemma]{lem:expected-dimension}
	For any fixed sketch matrix $\mA$, we have $\E_{\sC}\bigl[K(\sC)\bigr] \le 4 N$.
\end{lemma}

To prove \cref{lem:expected-dimension}, we need the following elementary lemma that upper bounds the number of positive eigenvalues of the difference of two PSD matrices by a trace.

\begin{lemma} \label[lemma]{lem:positive-eigenvalue-bound}
	For any matrices $\mM,\mN \succeq \mzero$ of the same dimension, we have
	\begin{align*}
		n_{+}(\mM-\mN) \le 2 \tr\bigl((\mM+\mN)^{\dagger} \mM\bigr).
	\end{align*}
\end{lemma}

\begin{proof}
	Write $\mW \coloneqq (\mM+\mN)^{\dagger/2} \, \mM \, (\mM+\mN)^{\dagger/2}$.
	By $\mM,\mN \succeq \mzero$, we have $\range(\mM) \subseteq \range(\mM+\mN)$.
	Since $(\mM+\mN)^{1/2} (\mM+\mN)^{\dagger/2}$ is the orthogonal projector onto $\range(\mM+\mN)$, this implies that $(\mM+\mN)^{1/2} \, \mW \, (\mM+\mN)^{1/2} = \mM$.
	Thus, we can write
	\begin{align}
		\mM - \mN = 2 \mM - (\mM + \mN) = (\mM+\mN)^{1/2} \bigl( 2 \mW - \mI \bigr) (\mM+\mN)^{1/2}. \label{eqn:whitening}
	\end{align}
	Note that $(\mM+\mN)^{1/2}$ is invertible on $\range(\mM+\mN)$ and both $\mM-\mN$ and $\mM$ are zero maps on $\range(\mM+\mN)^{\perp}$.
	Therefore, we have
	\begin{align*}
		n_{+}(\mM-\mN) & = n_{+}\bigl((\mM-\mN) |_{\range(\mM+\mN)}\bigr) \\
		& = n_{+}\Bigl(\bigl(2 \mW - \mI\bigr) \big|_{\range(\mM+\mN)}\Bigr) = n_{+}\bigl(2 \mW- \mI\bigr),
	\end{align*}
	where the second equality applies Sylvester's law of inertia on $\range(\mM+\mN)$ to \cref{eqn:whitening}.
	Thus, we have shown that $n_{+}(\mM-\mN)$ equals the number of eigenvalues of $\mW$ that are larger than $1/2$.
	As the sum of the eigenvalues of $\mW \succeq \mzero$ is $\tr(\mW) = \tr\bigl((\mM+\mN)^{\dagger} \mM\bigr)$, we conclude that $n_{+}(\mM-\mN) \le 2 \tr\bigl((\mM+\mN)^{\dagger} \mM\bigr)$, as desired.
\end{proof}

\begin{proof}[Proof of \cref{lem:expected-dimension}]
	We choose a uniform vertex $v \in \sR$ conditional on $\sC$ and aim to upper bound $\E\bigl[k_v(\sC)\bigr]$.
	Consider conditioning on the set $\sC' \coloneqq \sC \cup \{v\}$, so the vertex $v$ is a uniform vertex of $\sC'$, and the anchor set is $\sC' \setminus \{v\}$.
	Write $\mA_{\sC'} \coloneqq \bigl[\va_e: e \in \binom{\sC'}{2}\bigr]$.
	For each $u \in \sC'$, we can derive
	\begin{align*}
		k_u\bigl(\sC' \setminus \{u\}\bigr) & = n_{+}\Biggl(\sum_{w \in \sC' \setminus \{u\}} \va_{uw} \va_{uw}^{\top} - \sum_{e \in \binom{\sC' \setminus \{u\}}{2}} \va_e \va_e^{\top}\Biggr) \\
		& \le 2 \tr\biggl(\bigl(\mA_{\sC'} \mA_{\sC'}^{\top}\bigr)^{\dagger} \sum_{w \in \sC' \setminus \{u\}} \va_{uw} \va_{uw}^{\top}\biggr) \\
		& = 2 \sum_{w \in \sC' \setminus \{u\}} \tr\Bigl(\va_{uw}^{\top} \bigl(\mA_{\sC'} \mA_{\sC'}^{\top}\bigr)^{\dagger} \va_{uw}\Bigr) \\
		& = 2 \sum_{w \in \sC' \setminus \{u\}} \lev_{\va_{uw}}(\mA_{\sC'}),
	\end{align*}
	where the first line uses the definition of $k_u\bigl(\sC' \setminus \{u\}\bigr)$, the second applies \cref{lem:positive-eigenvalue-bound}, the third uses properties of trace, and the last uses the formula of leverage score.
	Thus, summing over $u \in \sC'$ gives
	\begin{align*}
		\sum_{u \in \sC'} k_u\bigl(\sC' \setminus \{u\}\bigr) \le 4 \sum_{e \in \binom{\sC'}{2}} \lev_{\va_{e}}(\mA_{\sC'}) = 4 \rank(\mA_{\sC'}) \le 4 N.
	\end{align*}
	Therefore, averaging over the uniform distinguished vertex $v \in \sC'$ and then over $\sC'$ yields $\E\bigl[k_v(\sC)\bigr] \le 4 N / (\ell b + 1)$.
	Since $v$ is uniform in $\sR$ conditional on $\sC$, we have $\E_{\sC}\bigl[K(\sC)\bigr] = \ell b \, \E\bigl[k_v(\sC)\bigr] \le 4 N$, as desired.
\end{proof}

\subsection{Proving the Lower Bound for Arbitrary Linear Sketches}

We are now ready to prove our main result, the sketching dimension lower bound of weighted spectral sparsification for arbitrary linear sketches.

\begin{proof}[Proof of \cref{thm:main}]
	Choose an arbitrary integer $t \ge 1$ and set $b \coloneqq \bigl(10^8 + 1\bigr) t + 1$, $\ell \coloneqq 10^7 b^4$, and $n \coloneqq 2 \ell b = \Theta\bigl(t^5\bigr)$.
	Suppose the sketch has dimension $N \le 10^{-12} n t$.
	Note that $n t = \Theta\bigl(n^{6/5}\bigr)$.
	The constructed inputs have positive-weight ratio $O(n^5)$, so by \cref{lem:anchored-reduction-to-leverage} it suffices to prove that, for every such fixed sketch matrix $\mA$,
	\begin{align*}
		\E_{\pi}\Bigl[ \min\Bigl\{ 1, \lev_{\va_q}\Bigl(\bigl(\mA \diag(\vw)\bigr)_{-q}\Bigr) \Bigr\} \Bigr] < \frac{1}{100}.
	\end{align*}

	Consider first revealing the anchor set $\sC$ and leaving the ring labeling uniform.
	By \cref{lem:expected-dimension} and Markov's inequality, we have
	\begin{align*}
		\Pr\nolimits_{\sC}\bigl\{K(\sC) > 10^{-9} \ell b t\bigr\} \le \frac{10^9 \E_{\sC}[K(\sC)]}{\ell b t} \le \frac{8 \cdot 10^9 N}{nt}.
	\end{align*}
	For every fixed $\sC$ with $K(\sC) \le 10^{-9} \ell b t$, \cref{lem:incidence-decomp} gives a fixed incidence sketch matrix $\mB$ of dimension $K(\sC)$ to which \cref{lem:endpoint-representation} applies under the uniform ring labeling.
	Let $\event$ and $\vf = (f_e)_{e \in \binom{\sR}{2} \setminus \{q\}}$ be the success event and coefficient vector supplied by \cref{lem:endpoint-representation} for each such $\sC$.
	Declare $\event$ false for all other anchor sets, and set $\vf = \vzero$ whenever $\event$ does not occur.
	Then $\Pr\{\event \mid \sC\} \ge 0.999$ whenever $K(\sC) \le 10^{-9} \ell b t$, and hence
	\begin{align*}
		1 - \Pr\{\event\} \le \frac{8 \cdot 10^9 N}{nt} + 10^{-3}.
	\end{align*}

	The coefficients supplied by \cref{lem:endpoint-representation} satisfy $f_e = 0$ whenever $\vb_e = \vzero$ and $\|\vf\|_2^2 \le 2 \ell b$.
	Hence, if $\event$ happens and $q$ is anchor-good, applying \cref{lem:lifting} gives coefficients $\vc = (c_e)_{e \in \binom{\sC}{2}}$ such that
	\begin{align*}
		\va_q & = \sum_{e \in \binom{\sR}{2} \setminus \{q\}} f_e \va_e + \mA_{\sC} \vc, \\
		\|\vc\|_2^2 & \le 6 \bigl(1 + \ell^2 b^2 \|\vf\|_2^2\bigr) \le 6 + 12 \ell^3 b^3.
	\end{align*}
	All anchor and within-block edges have base weight $w_{\mathrm{int}} = n^2$, so their total contribution to the squared coefficient norm for the scaled columns $w_e \va_e$ is at most
	\begin{align*}
		\frac{\|\vc\|_2^2 + \|\vf\|_2^2}{w_{\mathrm{int}}^2} \le \frac{6 + 12 \ell^3 b^3 + 2 \ell b}{16 \ell^4 b^4} \le \frac{1}{\ell b}.
	\end{align*}
	Thus, on this event, the definition of leverage score gives
	\begin{align*}
		\lev_{\va_q}\Bigl(\bigl(\mA \diag(\vw)\bigr)_{-q}\Bigr) \le \frac{1}{\ell b} + \sum_{0 \le i < j < \ell} \, \sum_{\substack{u \in \sB_i, v \in \sB_j, \\ uv \ne q}} \frac{f_{uv}^2}{w_{uv}^2}.
	\end{align*}

	For each fixed $\sC$ with $K(\sC) \le 10^{-9} \ell b t$, the cross-block estimate \eqref{eqn:incidence-cross-energy} applies conditionally on $\sC$, since $\mB$ is fixed, the ring labeling is uniform, and the cross-block base weights are unchanged.
	Since $\vf = \vzero$ for all other anchor sets, averaging over $\sC$ gives
	\begin{align*}
		\E_{\pi}\Biggl[\sum_{0 \le i < j < \ell} \, \sum_{\substack{u \in \sB_i, v \in \sB_j, \\ uv \ne q}} \frac{f_{uv}^2}{w_{uv}^2}\Biggr] \le \frac{6400 \, b^4}{\ell}.
	\end{align*}
	Restricting this nonnegative contribution to the event that $\event$ happens and $q$ is anchor-good can only decrease its expectation.
	If $\event$ fails or $q$ is not anchor-good, we use the trivial bound $1$ on the truncated leverage score.
	Recall from \cref{lem:anchor-good-prob} that the query edge $q$ is not anchor-good with probability at most $2N / \binom{\ell b+2}{2} \le 16 N / n^2 \le 16 N / (nt)$.
	Thus, by the union bound and the above bounds, we have
	\begin{align*}
		& \E_{\pi}\Bigl[ \min\Bigl\{ 1, \lev_{\va_q}\Bigl(\bigl(\mA \diag(\vw)\bigr)_{-q}\Bigr) \Bigr\} \Bigr] \\
		\le {} & \bigl(1 - \Pr\{\event\}\bigr) + \Pr\{q \text{ is not anchor-good}\} + \frac{1}{\ell b} + \frac{6400 \, b^4}{\ell} \\
		\le {} & \frac{8 \cdot 10^9 N}{nt} + 10^{-3} + \frac{16 N}{nt} + \frac{1}{\ell b} + \frac{6400 \, b^4}{\ell} < \frac{1}{100},
	\end{align*}
	where the last inequality follows from $N \le 10^{-12} nt$, $\ell = 10^7 b^4$, and $b \ge 10^8$.

	Finally, by padding the hard-instance graphs with isolated vertices, we can extend the lower bound to all sufficiently large $n$.
	This completes the proof of \cref{thm:main}.
\end{proof}

\paragraph{Choice of parameters and the exponent $6/5$.}
We take $b = \Theta(t)$ as in \cref{lem:endpoint-representation}, and $\ell = \Theta\bigl(b^4\bigr)$ with a sufficiently large constant to keep the cross-block contribution small.
These choices give $n = 2 \ell b = \Theta\bigl(t^5\bigr)$, so $n t = \Theta\bigl(n^{6/5}\bigr)$.
Since the proof excludes sketch dimensions up to a small constant multiple of $n t$, we obtain the lower bound $\Omega\bigl(n^{6/5}\bigr)$.

\section{Acknowledgments}

Part of this work was done during Mingji Yang's visit to the National University of Singapore, with partial support from the China Scholarship Council under Grant No. 202506360304.

\section{AI Methodology}

The proof of our result was initially found by GPT-6 Astra and then significantly simplified through extensive interactions with the model.
The authors have verified and rewritten the proof to improve readability and presentation.
The figures in this paper were prepared with the aid of GPT-6 Astra.

\printbibliography

\end{document}